\documentclass[nofootinbib,amsmath,amssymb,aps,prd,preprint]{revtex4-1}
\pdfoutput=1

\usepackage{graphicx}
\usepackage{natbib}
\usepackage{hyperref}

\newcommand{\der}{\mathrm{d}}
\newcommand{\mpl}{M_{\mathrm{P}}}

\begin{document}

\title{Dynamical Analysis of Attractor Behavior in Constant Roll Inflation}

\newcommand{\FIRSTAFF}{\affiliation{The Oskar Klein Centre for Cosmoparticle Physics,
	Department of Physics,
	Stockholm University,
	AlbaNova,
	10691 Stockholm,
	Sweden}}
\newcommand{\SECONDAFF}{\affiliation{Department of Physics,
	University at Buffalo, SUNY
	Buffalo,
	NY 14260
	USA}}
	
\author{Wei-Chen Lin}
\email{weichenl@buffalo.edu} 
\SECONDAFF
\author{Michael J. P. Morse}
\email{mjmorse3@buffalo.edu}
\SECONDAFF
\author{William H. Kinney}
\email{whkinney@buffalo.edu}
\SECONDAFF
\FIRSTAFF

\date{\today}

\begin{abstract}
There has been considerable recent interest in a new class of non-slow roll inflationary solutions known as \textit{constant roll} inflation. Constant roll solutions are a generalization of the ultra-slow roll (USR) solution, where the first Hubble slow roll parameter $\epsilon$ is small, but the second Hubble slow roll parameter $\eta$ is not. While it is known that the USR solutions represent dynamical transients, there has been some disagreement in literature about whether or not large-$\eta$ constant roll solutions are attractors or are also a class of transient solutions. In this paper we show that the large-$\eta$ constant roll solutions do in fact represent transient solutions by performing stability analysis on the exact analytic (large-$\eta$) constant roll solutions.   
\end{abstract}

\maketitle
\section{Introduction}
The constant roll inflation model was developed in recent work by Motohashi \textit{et. al.} \cite{Motohashi:2014ppa} as a generalization to the ultra-slow roll (USR) inflationary solution \cite{Kinney:1997ne,Tsamis:2003px,Kinney:2005vj}. Under the constant roll formalism, the second Hubble slow roll parameter $\eta$ is taken to be exactly constant, allowing the form of the Hubble parameter and potential to be analytically derived. Since it is also possible to analytically solve for the field, the dynamics are entirely determined and the solutions are potentially of great interest. Constant roll inflation and variants have been widely studied in the literature \cite{Tsamis:2003px,Kinney:1997ne,Martin:2012pe,Inoue:2001zt,Kinney:2005vj,Namjoo:2012aa,Huang:2013lda,Mooij:2015yka,Cicciarella:2017nls,Akhshik:2015nfa,Scacco:2015spa,Barenboim:2016mmw,Cai:2016ngx,Odintsov:2017yud,Grain:2017dqa,Odintsov:2017qpp,Bravo:2017wyw,Bravo:2017gct,Dimopoulos:2017ged,Nojiri:2017qvx,Motohashi:2017vdc,Odintsov:2017hbk,Oikonomou:2017xik,Cicciarella:2017nls,Awad:2017ign,Anguelova:2017djf,Salvio:2017oyf,Yi:2017mxs,Cai:2017bxr,Mohammadi:2018oku,Gao:2018tdb,Gao:2018cpp,Anguelova:2018ntr,Mohammadi:2018wfk,Karam:2017rpw,Morse:2018kda,Cruces:2018cvq,Mohammadi:2018zkf,Boisseau:2018rgy,Firouzjahi:2018vet,Matarrese:2018qqo,Pattison:2018bct,Ozsoy:2019slf,Leon:2019jsl,Gao:2019sbz}. 

In this paper, we briefly review the constant roll scenario and revisit the argument in Ref. \cite{Morse:2018kda} that when (large-$\eta$) constant roll solutions appear to be an attractor, it is due to a duality relation  $\eta \leftrightarrow (3-\eta)$. In Ref. \cite{Morse:2018kda} two of us (Kinney and Morse) presented a duality-based analytic argument that slow roll is the unique attractor solution in all cases. Here we present a direct numerical analysis of the stability of the analytic constant roll solution under the same duality. We show that the exact $\eta > 1.5$ constant roll solution is unstable to small perturbations, and that the solution relaxes (with time-dependent $\eta$) at late time to the dual attractor solution with $\tilde\eta = 3 - \eta < 1.5$. We also consider specifically the observationally viable solutions derived by Gao, {\it et al.} \cite{Gao:2019sbz}. We show that the large-$\eta$ solution is a transient, and that the attractor solution is the small-$\eta$ slow roll solution.\footnote{Such early-time transients may be of phenomenological interest, for example cases where the field is kinetically dominated prior to inflation can have suppressing effects on the low-$\ell$ multipoles in the CMB power spectrum, consistant with observed large scale anomalies \cite{Nicholson:2007by,Contaldi:2003zv}.}

This paper is structured as follows: In Section \ref{sec:HJ} we give a brief review of the constant roll formalism. The main body of the paper is in Section \ref{sec:Stability}, where we analyze the stability of large-$\eta$ constant roll solutions, and show that under perturbations $\eta$ evolves to the smaller of $\left\lbrace \eta,\tilde{\eta} = 3- \eta \right\rbrace $, and comment on the results of Ref. \cite{Gao:2019sbz}. Section \ref{sec:PhaseSpace} discusses stability of solutions in a phase space picture of the dynamics, and Section \ref{sec:Conclusion} presents conclusions.

\section{Constant Roll Inflation}
\label{sec:HJ}

In this work we consider cosmological inflation \cite{Starobinsky:1980te,Sato:1981ds,Sato:1980yn,Kazanas:1980tx,Guth:1980zm,Linde:1981mu,Albrecht:1982wi} driven by a single, minimally coupled canonical scalar field. The field dynamics are governed by the Friedmann equation and the equation of motion (EOM),
\begin{align}
\label{eq:Friedmann}
&H^2 = \frac{1}{3 M_P^2} \left[\frac{1}{2} \dot\phi^2 + V\left(\phi\right)\right], \\
\label{eq:inflEOM}
&\ddot\phi + 3 H \dot\phi + V'\left(\phi\right) = 0.
\end{align}
We assume no spatial curvature and take a metric of the Friedmann-Robertson-Walker form 
\begin{equation}
ds^2 = dt^2 - a^2\left(t\right) d{\bf x}^2,
\end{equation}
with the {\it Hubble parameter} $H$ defined as
\begin{equation}
H \equiv \left(\frac{\dot a}{a}\right).
\end{equation}
Inflation then consists of a period of accelerated expansion, $\ddot a > 0$, which corresponds to evolution dominated by the potential of the inflaton field, $\dot\phi^2 \ll V\left(\phi\right)$, so that
\begin{equation}
p \simeq - V\left(\phi\right) \simeq - \rho.
\end{equation}
If the evolution of the scalar field $\phi$ is monotonic, we can write the equations of motion (\ref{eq:inflEOM}) in the {\it Hamilton-Jacobi} form \cite{Muslimov:1990be,Salopek:1990jq,Lidsey:1995np},
\begin{align}
-2 \mpl^2 H' &= \dot{\phi},
\label{eq:FriedmannH'} \\
2[H'(\phi)]^2 - \frac{3}{\mpl^2}H^2(\phi) + \frac{1}{\mpl^4}V(\phi) &= 0.
\label{eq:HJ_eq}
\end{align}
We then define the familiar {\it slow roll parameters} in terms of the Hubble parameter $H\left(\phi\right)$ as:
\begin{eqnarray}
\label{eq:defSR}
&&\epsilon \equiv \frac{\dot\phi^2}{2 \mpl^2 H^2} = 2\mpl^2 \left(\frac{H'\left(\phi\right)}{H\left(\phi\right)}\right)^2,\cr
&&\eta \equiv \frac{-\ddot \phi}{H \dot\phi} = 2 \mpl^2 \frac{H''\left(\phi\right)}{H\left(\phi\right)},
\end{eqnarray}
where a prime denotes differentiation with respect to the ``clock'' field $\phi$. Inflation, in the general sense of accelerated expansion, corresponds to $\epsilon < 1$. Typical inflationary dynamics obey the {\it slow roll} conditions,
\begin{equation}
\epsilon , \eta \ll 1,
\end{equation}
resulting in a sufficiently long period of quasi-de Sitter expansion and a nearly scale-invariant spectrum of density perturbations. 

There has been considerable recent interest in a class of non-slow roll solutions, {\it constant roll} inflation \cite{Tsamis:2003px,Kinney:1997ne,Martin:2012pe,Inoue:2001zt,Kinney:2005vj,Namjoo:2012aa,Huang:2013lda,Mooij:2015yka,Cicciarella:2017nls,Akhshik:2015nfa,Scacco:2015spa,Barenboim:2016mmw,Cai:2016ngx,Odintsov:2017yud,Grain:2017dqa,Odintsov:2017qpp,Bravo:2017wyw,Bravo:2017gct,Dimopoulos:2017ged,Nojiri:2017qvx,Motohashi:2017vdc,Odintsov:2017hbk,Oikonomou:2017xik,Cicciarella:2017nls,Awad:2017ign,Anguelova:2017djf,Salvio:2017oyf,Yi:2017mxs,Cai:2017bxr,Mohammadi:2018oku,Gao:2018tdb,Gao:2018cpp,Anguelova:2018ntr,Mohammadi:2018wfk,Karam:2017rpw,Morse:2018kda,Passaglia:2018ixg,Gao:2019sbz}, which are defined by constant slow-roll parameter 
\begin{equation}
\eta \equiv \bar\eta = \mathrm{const.}
\end{equation}
Elsewhere in the literature ({\it e.g.} Ref. \cite{Motohashi:2014ppa}), this constant is defined as 
\begin{equation}
\alpha = \bar\eta - 3.
\end{equation}
In general the constant $\bar{\eta}$ does not need to be small and can be larger than unity.
Applying the constant roll condition to the definition of $\eta$ (\ref{eq:defSR}) results in a simple differential equation for $H\left(\phi\right)$,
\begin{align}
H'' &= \frac{ \bar\eta}{2 \mpl^2}H,
\end{align}
with general solution 
\begin{align}
H(\phi) &= A \exp\left( \sqrt{\frac{\bar\eta}{2}} \frac{\phi}{\mpl}\right) \nonumber \\ &+ B \exp\left( -\sqrt{\frac{\bar\eta}{2}} \frac{\phi}{\mpl}\right).
\label{eqn:H_General_-a}
\end{align}
Since the form of the potential is fully determined through $H$ and $H'$ by Eq. (\ref{eq:HJ_eq}), it is trivial to write the general form of the potential
\begin{eqnarray}
V(\phi) = &&(3-\bar\eta) A^2 \exp\left(2 \sqrt{ \frac{\bar\eta}{2}} \frac{\phi}{\mpl}\right)  \nonumber \\ &&  + (3-\bar\eta) B^2 \exp\left( -2 \sqrt{\frac{\bar\eta}{2}} \frac{\phi}{\mpl}\right)\cr \nonumber \\
&&+ 2(\bar\eta)AB + 6AB. 
\end{eqnarray}
For suitable choice of boundary condition the potential and Hubble parameter can be written in the form \cite{Motohashi:2014ppa}
\begin{align}
\label{eq:V_eta>epsilon}
V(\phi) = 3 \mpl^2 H_0^2 \left(1 + \frac{1}{3} (3-\bar\eta) \sinh ^2\left(\sqrt{\frac{\bar\eta}{2}}\frac{\phi}{\mpl}\right)\right), \\
H(\phi) = H_0\cosh\left( \sqrt{\frac{\bar\eta}{2}} \frac{\phi}{\mpl}\right) .
\label{eq:V_hybrid}
\end{align}
The resulting potential has different qualitative form depending on the value of $\bar\eta$. For $0 < \bar\eta < 3$ the potential is a convex function with a minimum at $\phi = 0$ with $V\left(\phi = 0\right) = 3 M_P^2 H_0^2$ corresponding to a ``hybrid''-type inflation model where inflation continues indefinitely in the absence of an auxiliary field. For $\bar\eta > 3$ the potential is a concave function corresponding to a ``hilltop''-type inflation model where inflation takes place near the maximum of the potential, and ends as the potential steepens away from the maximum. 
In the case of a negative $\bar\eta$ we can write the potential as\footnote{We note that our earlier work \cite{Morse:2018kda} incorrectly identifies both the $\bar\eta<0$ and $\bar\eta>3$ concave potentials as $\sin^2(\phi)$ potentials such as Eq. (\ref{eq:V_eta_negative}). The $\bar\eta>3$ constant roll potentials are in fact of a $\sinh^2(\phi)$ form (\ref{eq:V_eta>epsilon}.  The duality $\bar\eta \rightarrow \bar\eta - 3$ then maps between the $\sinh(\phi)^2$ hilltop potential and the $\sin(\phi)^2$ case, which are equivalent \textit{only} in the small field limit. This does not affect the discussion or conclusions of Ref. \cite{Morse:2018kda}.}
\begin{align}
\label{eq:V_eta_negative}
V(\phi) = 3 \mpl^2 H_0^2\left(1 - \frac{1}{3} (3-\bar\eta) \sin ^2\left(\sqrt{\frac{-\bar\eta}{2}} \frac{ \phi }{\mpl}\right)\right).
\end{align}
In this case the potential is again a concave function corresponding to ``hilltop''-type inflation models. 
\begin{figure}
\includegraphics[width = 0.8\textwidth]{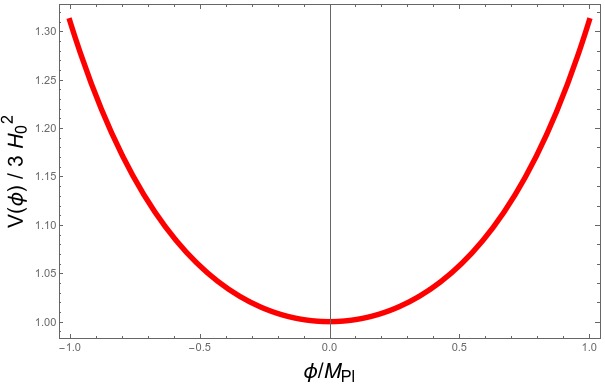}
\caption{The form of the convex potential, for $\bar\eta = 2.5$.}
\label{fig:ConvexPotential}
\end{figure}
\begin{figure}
\includegraphics[width = 0.8\textwidth]{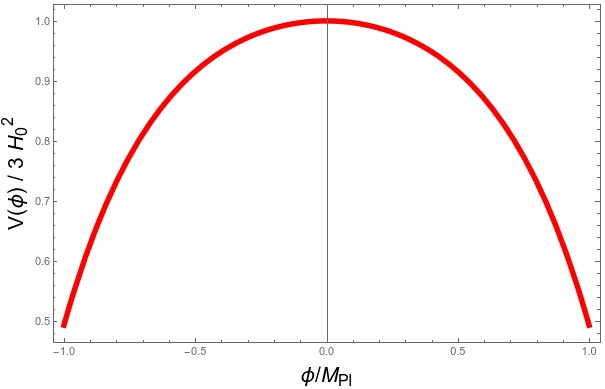}
\caption{The form of the concave potential, for $\bar\eta = 3.5$.}
\label{fig:ConcavePotential}
\end{figure}

Equation (\ref{eq:FriedmannH'}) can be used to determine the form of the dynamical solution $\phi(t)$. For $3>\bar\eta>0$, the constant roll analytic solution has the form
\begin{align}
\frac{\phi(t)}{\mpl} = 2\sqrt{\frac{2}{\bar\eta}} \mathrm{arctanh}\left( e^{- \bar{\eta}H_0t}\right).
\label{eq:hybridfield}
\end{align}
A complete discussion of parameters and associated solutions can be found in \cite{Motohashi:2014ppa}. A key point of the analysis here is that the solution (\ref{eq:hybridfield}) is a {\it particular} solution to the equation of motion for the field: the value of the slow roll parameter $\eta$ for a general field solution $\phi\left(t\right)$ can be time-dependent and not necessarily equal to the constant $\bar\eta$.  In this paper we show that the constant roll solution (\ref{eq:hybridfield})  with $\bar\eta > 1.5$ is in dynamically unstable with the generic late-time attractor being the dual $\eta = 3 - \bar\eta$. When $\bar{\eta} > 2$ the late-time attractor will be slow-roll with $\eta<1$.

\section{Stability analysis of constant roll inflation}
\label{sec:Stability}

In this section we consider perturbations to the exact constant roll solutions, using the accompanying code \cite{michael_j_p_morse_2019_3066362}. We show that the solution with $\eta < 1.5$ is the unique attractor, corresponding to slow roll for $\eta < 1$. In order to check if the constant roll analytic solution is a dynamical attractor, we look at perturbations of the initial condition $\dot{\phi}$ around the constant roll analytic solution as is done in \cite{Motohashi:2014ppa} (see Fig. 2 within). However, in Ref. \cite{Motohashi:2014ppa} only the phase portrait $(\phi,\dot\phi)$ is examined to assess stability. In Ref. \cite{Morse:2018kda} we argue that the apparent attractor behavior is an artifact of a parameter duality which relates the $\eta = \bar\eta > 1.5$ large-$\eta$ constant roll solution to a corresponding small-$\eta$ solution under the duality $\eta = 3 - \bar\eta$, and that the unique attractor is always slow roll when it exists. In this paper we examine the attractor behavior of constant roll and the associated dual slow roll solution using a direct dynamical approach by considering the time evolution of $\eta$ under small perturbations about the analytic constant roll solution. We find that under perturbations, dynamics which is initially $\eta = \bar\eta > 1.5$ generically relaxes after a few e-folds to the attractor dual solution, $\eta = 3 - \bar\eta$. We consider two cases: First, in the case $0<\eta<3$, the exact constant roll solution corresponds a field rolling down on a convex (hybrid-type) potential. Second, we consider $\eta>3$, where the exact solution corresponds to a field climbing {\it up} a concave (hilltop-type) potential. We use the full field equations to evolve the system
\begin{align}
\ddot{\phi} + 3H \dot{\phi} + \frac{\der V}{\der \phi} &= 0 ,\\
3 \mpl^2 H^2 &= \frac{\dot{\phi}^2}{2} + V.
\end{align}
Using the full dynamical equations instead Hamilton-Jacobi Formalism allows us to directly describe situations where the field evolution is non-monotonic. We work in dimensionless variables: $\tilde{\phi} = \frac{\phi}{\mpl}$ and $\tilde{t} = H_0 t$. Since each term in the field EOM and Friedmann equation are proportional to $H_0^2$ and $\mpl$, the dynamics will be scale independent and we are free to set $\mpl$, $H_0$ to unity in the analysis.

\subsection{Convex Potential}
\label{subsec:Hybrid}
In this section we consider convex potentials, $V''(\phi) > 0$ and $V(\phi = 0) \equiv V_0 \neq 0$, so that the asymptotic behavior of the field is to come to rest at the minimum of the potential (taken to be at the origin), and inflation continues forever in the absence of an auxiliary field to end inflation via the so-called ``hybrid'' mechanism  \cite{Linde:1993cn}.
Such models are currently of little phenomenological interest, since even in the slow roll limit $0 < \eta \ll 1$, the power spectrum is ``blue'', $n_S > 1$, which is strongly inconsistent with data. Such solutions are nonetheless of dynamical interest, and we include them here for completeness. We consider the phenomenologically viable ``hilltop''-type, or concave potentials, in Sec. \ref{subsec:Concave_potential}.

The constant roll analytic solution of the field (\ref{eq:hybridfield}) will asymptote to rest at the bottom of the potential as time goes to infinity, \textit{i.e.} $\dot\phi \rightarrow 0$. However, since we have a relation of the field acceleration to the field velocity,
\begin{align}
\frac{-\ddot{\phi}}{\dot{\phi} H } = \eta = \bar{\eta},
\end{align}
the exact form of the asymptotic behavior is uniquely determined by the value of $\eta$ so that the $\eta = \bar\eta > 2$ constant roll solution and the dual $\eta = 3 - \bar\eta < 1$ solution (slow roll) are dynamically distinct.

In Fig. \ref{fig:eta_n=2.5} we have taken $\bar\eta = 2.5$ in the potential (\ref{eq:V_hybrid}), such that the constant roll analytic solution is inherently non-slow roll. We perturb $\dot{\bar{\phi}}$, the constant roll analytic solution, in $1 \%$ intervals in order to probe the attractor behavior, 
\begin{align}
\delta\dot{\phi} =  \frac{\dot{\bar{\phi}}}{100}.
\end{align}

\begin{figure}
\includegraphics[width = 0.8\textwidth]{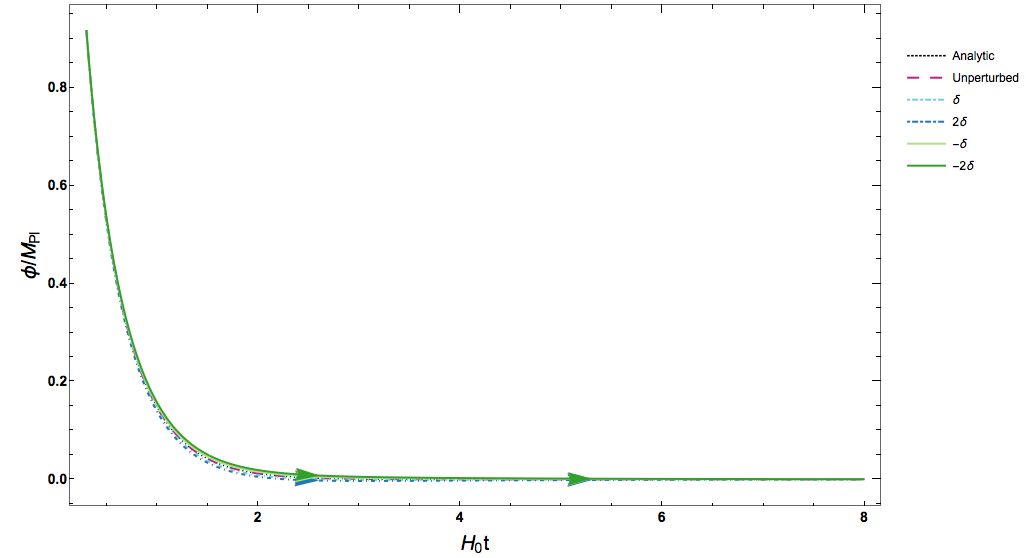}
\caption{Evolution of $\phi$ on the convex potential for various trajectories perturbed about the $\eta = \bar\eta = 2.5$ constant roll solution, labeled ``Analytic''.}
\label{fig:phi_n=2.5}
\end{figure}

\begin{figure}
\includegraphics[width = 0.8\textwidth]{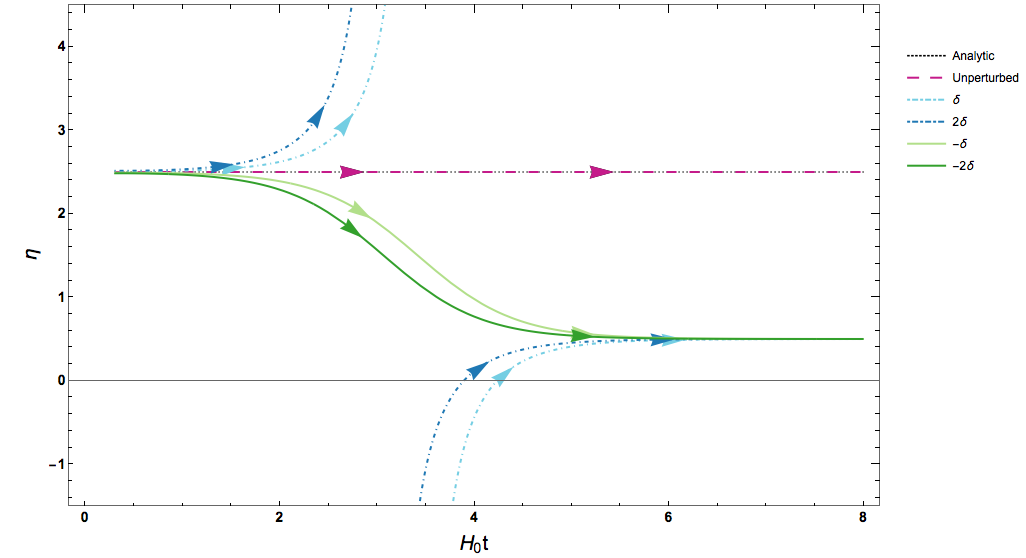}
\caption{Evolution of $\eta$ on the convex potential for various trajectories perturbed about the $\eta = \bar\eta = 2.5$ constant roll solution, labeled ``Analytic''. The late-time attractor solution is the dual slow roll solution with $\eta = 3 - \bar\eta = 0.5$.}
\label{fig:eta_n=2.5}
\end{figure}

\begin{figure}
\includegraphics[width = 0.8\textwidth]{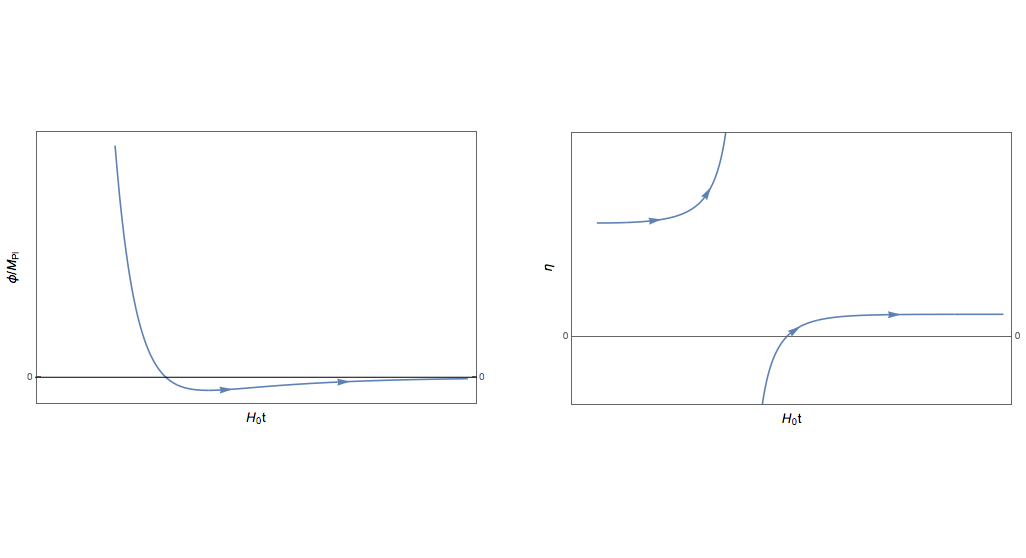}
\caption{Evolution of $\phi$ and $\eta$ in a convex potential for an increased initial field speed, with $\bar\eta = 2.5$. Here the field ``overshoots'' the minimum, turns around, and relaxes asymptotically to the minimum along a slow roll trajectory.}
\label{fig:ssos_n=2.5}
\end{figure}

\begin{figure}
\includegraphics[width = 0.8\textwidth]{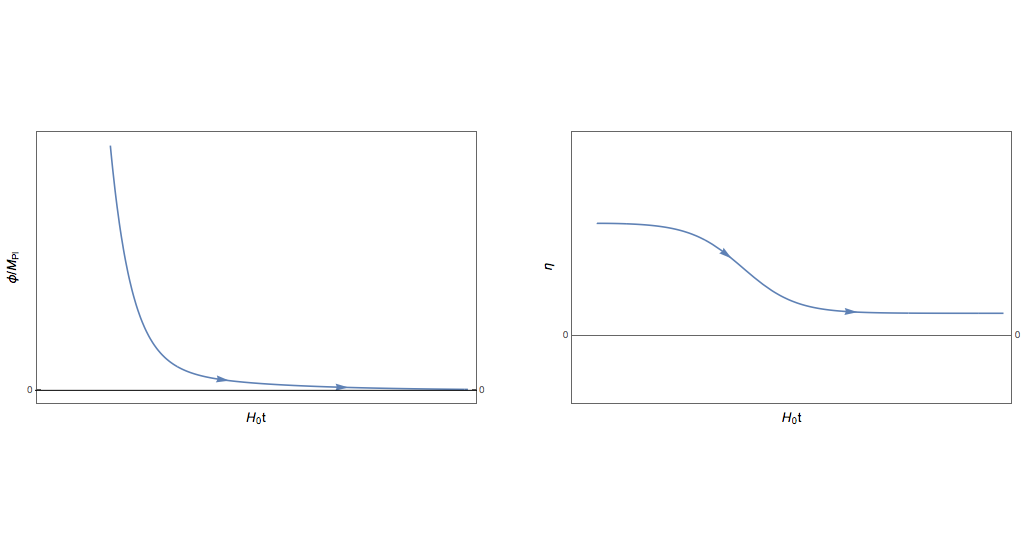}
\caption{Evolution of $\phi$ and $\eta$ in a convex potential for an decreased initial field speed, with $\bar\eta = 2.5$. Here the field relaxes directly to the slow roll attractor solution, and asymptotically approaches the minimum of the potential.}
\label{fig:ssus_n=2.5}
\end{figure}
A positive kick, $\delta > 0$, will increase the field speed causing the field to ``overshoot'' the potential minimum and roll a small distance {\it up} the opposite side of the potential, then relax back to the minimum along a slow roll trajectory. In this case we have a turning point where $\dot{\phi}=0$, and $\eta$ displays a discontinuity. A side-by-side plot of the dynamics of $\phi$ and $\eta$ for this case can be seen in Fig. \ref{fig:ssos_n=2.5}. For a negative kick, $\delta < 0$, the field speed is decreased and the field relaxes to a the slow-roll solution directly, and asymptotically evolves to rest at the origin. A side-by-side plot of the dynamics of $\phi$ and $\eta$ for this case can be seen in Fig. \ref{fig:ssus_n=2.5}. However, for both the positive and negative kick case, the attractor solution is not the analytic constant roll large-$\eta$ solution, but rather solution in the slow roll regime where $\tilde{\eta} = 3- \eta$ under the duality found in \cite{Kinney:1997ne,Tsamis:2003px,Morse:2018kda}. For Fig. \ref{fig:eta_n=2.5} this corresponds to $\tilde{\eta} = 0.5$ because the large-$\eta$ constant roll was chosen to be $\eta= 2.5$. Therefore, the slow roll solution is the attractor solution, a conclusion which is not confined to this particular choice of $\bar\eta$.

\begin{figure}
\includegraphics[width = 0.8\textwidth]{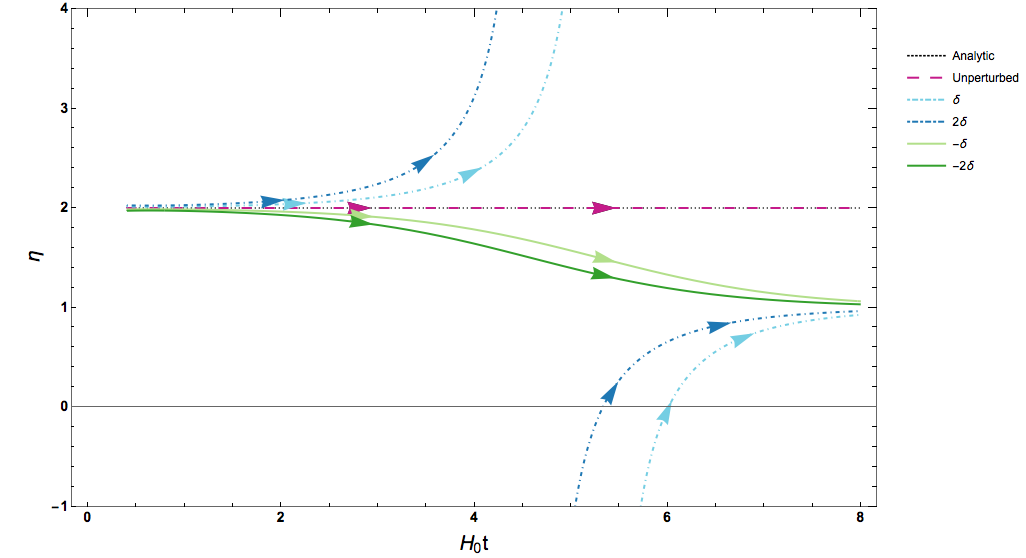}
\caption{Evolution of $\eta$ on convex potential with no slow roll, with an unstable solution $\eta = \bar\eta = 2.0$, and an attractor solution with $\eta = 3 - \bar\eta = 1.0$. }
\label{fig:eta_n=2}
\end{figure}

It is interesting to note that for $1<\bar\eta<2$ there is no slow roll solution, because the dual relation takes a larger $ 2 > \eta = \bar\eta > 1.5$ solution to a smaller $1.5>\eta = 3 - \bar\eta > 1$ solution.  This ``self-dual'' region will correspond to blue tilt in the scalar spectral index \cite{Morse:2018kda}, and is uninteresting from a phenomenological standpoint. A plot of this can be seen in Fig. \ref{fig:eta_n=2}.

\subsection{Concave Potential}
\label{subsec:Concave_potential}
We perform the same analysis on $\bar\eta>3$ potentials,
\begin{equation}
V(\phi) = 3 \mpl^2 H_0^2 \left(1 + \frac{1}{3} (3-\bar\eta) \sinh ^2\left(\frac{\sqrt{\bar\eta} }{\sqrt{2}}\frac{\phi}{\mpl}\right)\right),
\end{equation}
with field evolution
\begin{align}
\frac{\phi(t)}{\mpl} = 2\sqrt{\frac{2}{\bar\eta}} \mathrm{arctanh}\left( e^{- \bar{\eta}H_0t}\right).
\end{align}
The solutions with $\bar\eta > 3$ are of phenomenological interest since they are of the ``hilltop'' type which is compatible with current constraints. Note that the analytic constant-roll equation in this case corresponds to the field rolling {\it up} a concave potential and asymptotically coming exactly to rest at the maximum of the potential, taken to be the origin. (This is in contrast to the slow roll solution, for which the field approaches rest at the maximum as $t \rightarrow -\infty$.) Without some instability mechanism such as quantum perturbations, this family of background solutions leads classically to eternal inflation, since the field asymptotes to rest on top of the potential. It is found in \cite{Gao:2019sbz} that in this regime $\dot{\epsilon}<0$, and these solutions correspond to phenomenologically interesting scenarios, because the scalar spectral index and tensor to scalar ratio are within observable bounds \cite{Akrami:2018odb,Ade:2018gkx}. However, in such a circumstance it is necessary induce a mechanism to end inflation after 60 e-folds. This leads to an inherent ambiguity in the predictions for observables, which depend in a nontrivial way on the field value at the end of inflation. We show here that such background solutions, while valid, are in any case not stable under perturbations, and that the generic attractor solution is the dual slow roll case with $\eta \rightarrow 3- \bar\eta$, and field rolling {\it down} the potential. 

\begin{figure}
\includegraphics[width = 0.8\textwidth]{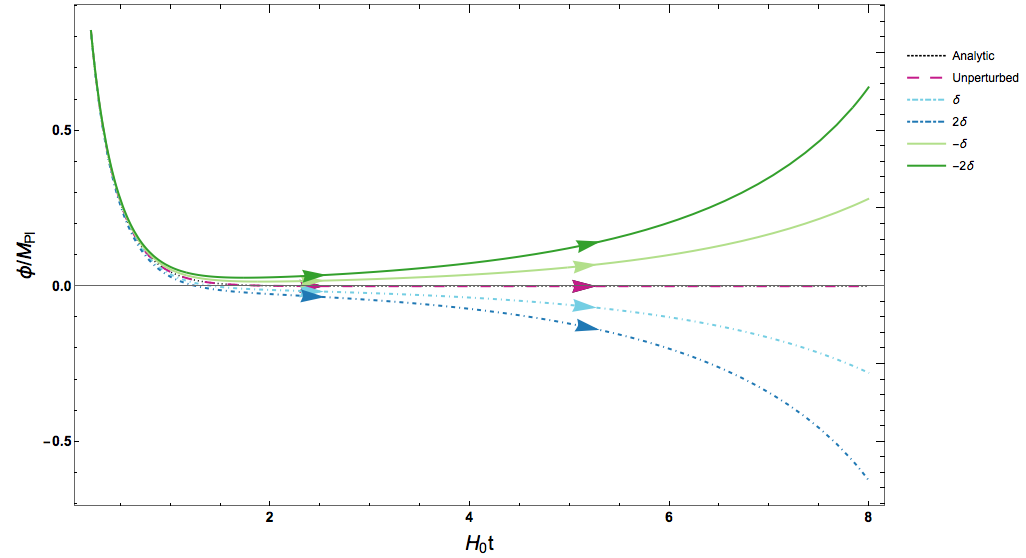}
\caption{Evolution of $\phi$ on the concave potential for various trajectories perturbed about the constant roll solution, labeled ``Analytic'', for $\bar\eta = 3.5$.}
\label{fig:phi_n=3.5}
\end{figure}

\begin{figure}
\includegraphics[width = 0.8\textwidth]{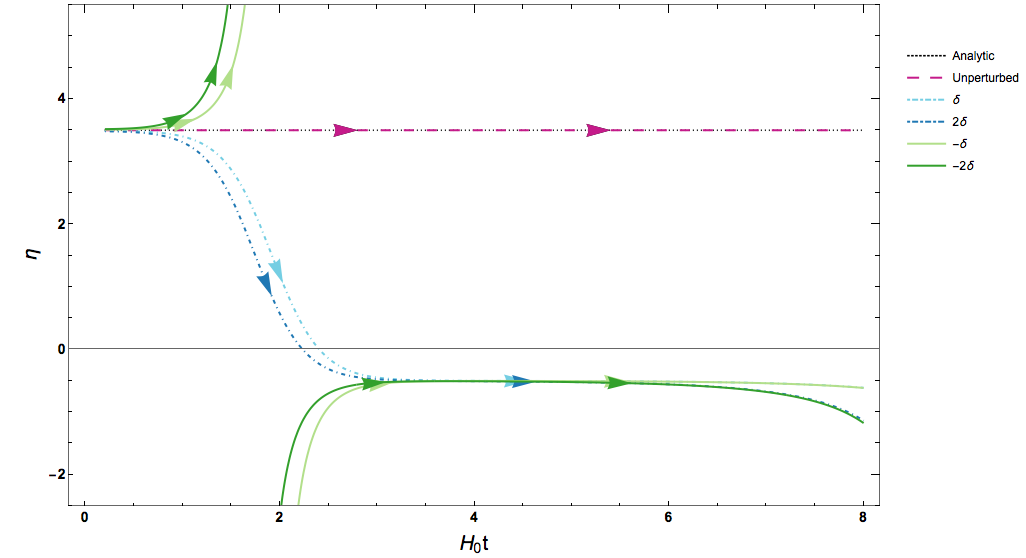}
\caption{Evolution of $\eta$ on the concave potential for various trajectories perturbed about the constant roll solution, labeled ``Analytic'', for $\bar\eta = 3.5$.}
\label{fig:eta_n=3.5}
\end{figure}

The methodology we use here is the same as in the convex potential case. We take small perturbations to the field velocity ($1\%$) and show that the constant roll solution is unstable. We choose $\bar\eta = 3.5$ and plot the analytic, numeric, and perturbed evolution in Fig. \ref{fig:phi_n=3.5} and Fig. \ref{fig:eta_n=3.5}. While the dual slow-roll solution $\eta = (3 - \bar\eta) = -0.5$ is not a good fit to data, this parameter choice makes the qualitative properties of the numerical solutions more evident. We consider the phenomenologically viable case $\left\vert \eta\right\vert \ll 1$ below. 

\begin{figure}
\includegraphics[width = 0.8\textwidth]{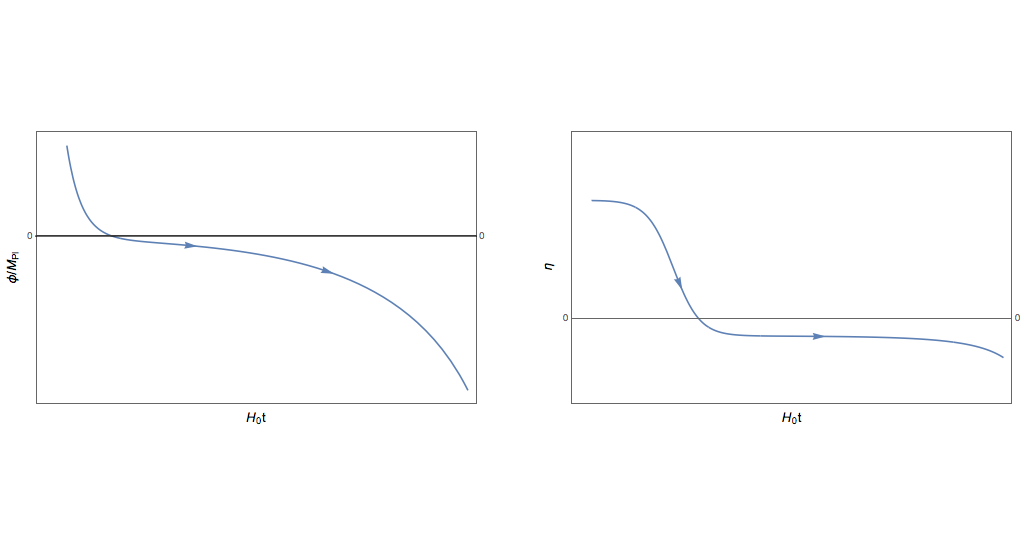}
\caption{Evolution of $\phi$ and $\eta$ in a concave potential for an increased initial field speed, with $\bar\eta = 3.5$. In this case, the field evolves over the maximum of the potential, and rolls down the far side along a slow roll trajectory.}
\label{fig:ssos_n=3.5}
\end{figure}

\begin{figure}
\includegraphics[width = 0.8\textwidth]{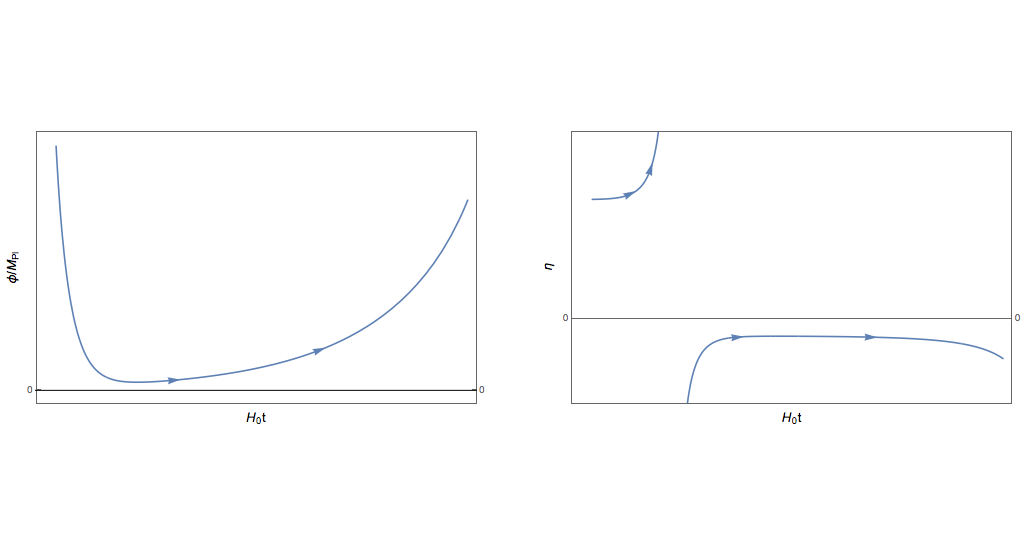}
\caption{Evolution of $\phi$ and $\eta$ in a concave potential for an decreased initial field speed, with $\bar\eta = 3.5$. In this case, the field turns around short of the potential maximum, and evolves in the opposite direction along a slow roll trajectory.}
\label{fig:ssus_n=3.5}
\end{figure}

It is then clear that the $\bar\eta > 3$ analytic constant roll solution is not stable under perturbations. There are again two qualitatively different regimes, depending on the sign of the perturbation $\delta$: When a positive kick is applied, the field speed is increased and the field will roll over the top of the hill, at which point it will roll down the far side of the potential maximum in the slow roll regime. A side-by-side plot of the dynamics of $\phi$ and $\eta$ for this case can be seen in Fig. \ref{fig:ssos_n=3.5}. Conversely, when a negative kick is applied to the system, the field will come to rest before it reaches the maximum of the potential, at which point $\dot\phi = 0$ and $\eta$ is discontinuous. The field will then begin to roll back the direction from which it came, down the hill in the slow roll regime.  A side-by-side plot of the dynamics of $\phi$ and $\eta$ for this case can be seen in Fig. \ref{fig:ssus_n=3.5}. Again, the attractor solution is not the constant roll analytic solution, $\eta = \bar\eta > 3$, but the solution in the slow roll regime under the dual relation $\eta = 3- \bar\eta$. It is likewise intuitively clear that the constant roll solution cannot be stable, since the initial conditions must be fine-tuned to cause the field to roll and asymptote to rest at the top of the hill. 

The reason for transitioning from the larger $ \eta = \bar\eta > 3$ to the smaller value $\eta = (3-\bar\eta) < 0 $ can be clearly seen by considering the field evolution near the maximum of the potential. As the field value decreases, the field evolves from a non-linear region to a linear region in which the dynamics of the field is governed by the linear differential equation,   
\begin{equation}
\ddot{\phi} + 3H_{0} \dot{\phi} + H_{0}^{2}\bar\eta(3-\bar\eta)\phi \approx 0,
\end{equation}
which is invariant under the duality $ \bar\eta \rightarrow 3- \bar\eta  $. The general solution of this linear differential equation is a linear combination of two special solutions \cite{Kinney:1997ne,Kinney:2005vj},
\begin{equation}
\phi(t)=A\exp[-H_{0}\bar\eta t]+B\exp[-H_{0} (3-\bar\eta) t]. 
\label{eq:smallFieldSolution}
\end{equation}
The analytic constant roll solution Eq. (\ref{eq:hybridfield}) corresponds to the particular boundary condition $B = 0$. For a general solution, $\eta = \eta(t)$ is time-dependent, and the $B \neq 0$ term always dominates at late time, and is therefore the unique attractor solution. This is clear if we give a small perturbation to the initial field velocity, $ \dot{\phi}(\phi_i) \rightarrow \dot{\phi}(\phi_i) + \delta $: although the field still evolves according to the Friedmann equation and the EOM of the field Eqs. (\ref{eq:Friedmann}, \ref{eq:inflEOM}), it no longer follows the evolution of the analytic $ \eta = \bar\eta = \mathrm{const.}$ solution. When the field enters the linear regime, perturbations to the constant-$ \eta $ solution are exactly described by the second special solution $ B\exp[-H_{0} (3-\eta) t]$ where $ B $ is the measure of the deviation from the analytic solution, and the field relaxes to the attractor. Unlike in the case of the convex potential, the field eventually evolves out of the small-field, slow roll limit as it rolls down the hill and inflation ends. Once out of the small-field limit, Eq. (\ref{eq:smallFieldSolution}) is no longer a good approximation to any general solution, and the system will evolve away from $\eta = \mathrm{const.}$ as seen in Fig. \ref{fig:eta_n=3.5} when $t \approx 7$.

There exists a second class of concave potentials on which constant-$\eta$ solutions exist. These potentials correspond to the choice of $\bar{\eta}<0$, and are of the form
\begin{equation}
V(\phi) = 3 \mpl^2 H_0^2 \left(1 - \frac{1}{3} (3-\bar\eta) \sin ^2\left(\frac{\sqrt{-\bar{\eta}} }{\sqrt{2}}\frac{\phi}{\mpl}\right)\right),
\end{equation}
with field evolution
\begin{align}
\frac{\phi(t)}{\mpl} = 2 \sqrt{\frac{2}{-\bar\eta}} \arctan \left( e^{-\bar{\eta}H_0 t} \right).
\label{eq:phi_negative_n}
\end{align}
While it may not be readily apparent, our time domain is restricted 
to $t \in (-\infty,0)$. To see this we may look at the form of the Hubble parameter as a function of time
\begin{align}
H(t) = - H_0 \tanh \left( -\bar{\eta}H_0 t \right).
\end{align}
Therefore, the analytic constant roll solution in this case corresponds to a field sitting at the top of the potential at $t \rightarrow - \infty$ and rolling \textit{down} as $t \rightarrow 0$, which is exactly the usual slow roll solution. When the field is near the top of the potential the solution is in the linear limit and Eq. (\ref{eq:phi_negative_n}) has the small field form of
\begin{align}
\frac{\phi(t)}{\mpl} \propto e^{ -\bar{\eta} H_{0} t}.
\end{align}
Here $\bar{\eta}$ is the smaller of the set $ \left\lbrace \bar{\eta},3-\bar{\eta}\right\rbrace$ as $\bar{\eta}<0$, therefore the analytic constant roll solution necessarily corresponds to the slow roll attractor.

The existence of the duality $\eta \rightarrow 3 - \eta$ does {\it not}, however, mean that the potentials (\ref{eq:V_eta>epsilon}) and (\ref{eq:V_eta_negative}) are identical, which they are clearly not, or that they make the same prediction for CMB observables. The potentials are identical only in the limit of $\phi / \mpl \ll 1$, where
\begin{equation}
V\left(\phi\right) \rightarrow 3 \mpl^2 H_0^2 \left[ 1 + \frac{1}{2}  \frac{\bar\eta \left(3 -\bar\eta\right)}{3} \left(\frac{\phi}{\mpl}\right)^2 + \cdots\right],
\end{equation}
which is the same region for which the perturbation equation becomes self-dual, and which is also the late-time limit in the convex (hybrid) case. In the concave (hilltop) case, slow roll breaks down entirely in the late-time limit, and {\it neither} the $\bar\eta = \mathrm{const.} > 3$ or the dual $\bar\eta = \mathrm{const.} < 0$ solution is valid. This means that the endpoint of inflation, when $\epsilon = 1$, is completely different in the two cases. For example, in the phenomenologically interesting case $\bar\eta = 3.0115$, which will be considered next, the endpoint of inflation on the potential (\ref{eq:V_eta>epsilon}) is at $\phi/\mpl \simeq 2.4$, while in the dual case $\bar\eta = -0.0115$ on the potential (\ref{eq:V_eta_negative}), the endpoint of inflation is at $\phi / \mpl \simeq 18.9$. A point of confusion which has arisen in the literature is the difference between the arbitrary constant $\bar\eta$ used to define the potentials (\ref{eq:V_eta>epsilon}) and (\ref{eq:V_eta_negative}), and the {\it dynamical} $\eta\left(t\right)$ which corresponds to the actual field evolution, which need not be equal to the $\bar\eta$ used to define the potential, and also need not be constant for a general dynamical solution of the field equation.

\begin{figure}
\includegraphics[width = 0.8\textwidth]{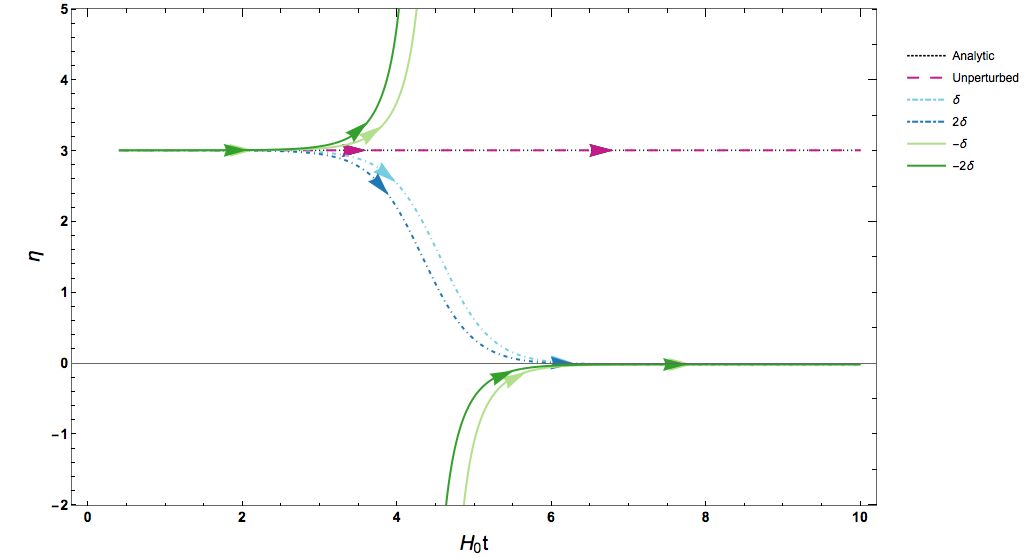}
\caption{Evolution of $\eta$ starting from $\epsilon = 1$ for $\eta = 3.0115$. }
\label{fig:eta_n=3.0115}
\end{figure}

To make contact with phenomenology, we apply our results to the almost flat case considered in Gao, {\it et al.}, \cite{Gao:2019sbz}, which corresponds to the phenomenologically interesting slow roll case in the dual limit, with $\eta \ll 1$. We again analyze stability of the $\bar\eta > 3$ solution by introducing perturbations of the initial condition $\dot{\phi}$ around the analytic constant roll solution. For numerical reasons, we take much smaller perturbations, on the order of $0.1\%$,
\begin{align}
\delta\dot{\phi} =  \frac{\dot{\bar{\phi}}}{1000}.
\end{align}
The result for the observationally favored case of $\eta = 3.0115$ is shown in Fig. \ref{fig:eta_n=3.0115}. The results are similar for the upper bound of $\eta = 3.024$ from Ref. \cite{Gao:2019sbz}.  
\begin{figure}
\includegraphics[width = 0.8\textwidth]{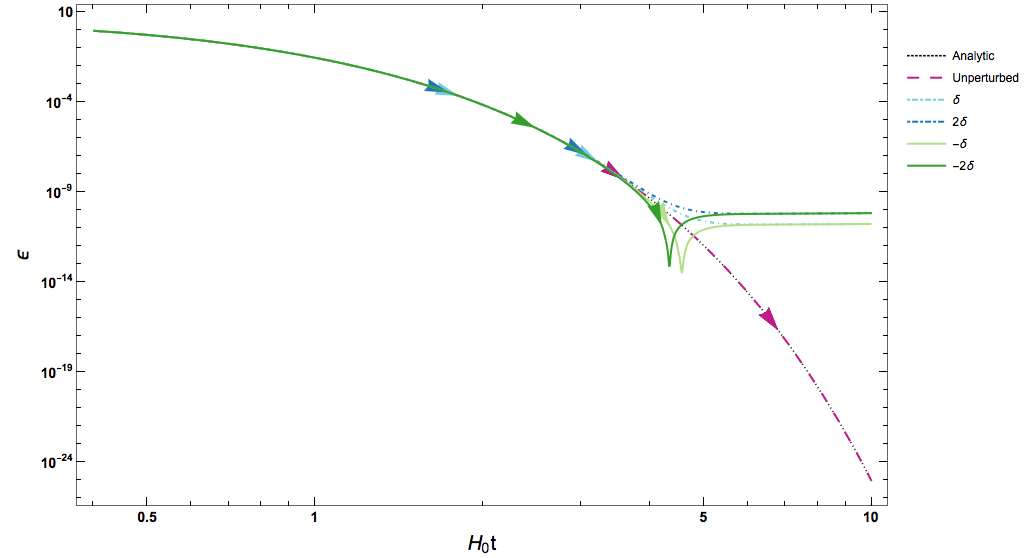}
\caption{Evolution of $\epsilon$ for $\eta = 3.0115$. Note that $H_0 t$ is to a good approximation the number of e-folds $N$, showing that in the slow roll phase, $\epsilon$ remains of $O\left(10^{-10}\right)$ for more than a hundred e-folds. }
\label{fig:epsilon_n=3.0115}
\end{figure}
We agree with Gao, {\it et al.},  that when $\epsilon$ becomes large the duality breaks down, because the first slow roll parameter $\epsilon \sim \left\vert \eta\right\vert$ and cannot be neglected in the expression for the scalar spectral index $n_S$, leading to different results in the dual large-$\eta$ and small-$\eta$ cases. This result was mentioned in \cite{Morse:2018kda} as well. The evolution of $\epsilon$ for the case $\bar\eta = 3.0115$ is shown in Fig. \ref{fig:epsilon_n=3.0115}. Note that the numerical solution is evaluated starting at the {\it onset} of inflation for the field rolling up the inverted potential, {\it i.e.} when $\epsilon = 1$. After the onset of inflation, the first slow roll parameter decreases exponentially during the $\eta > 3$ phase, reaching of order $10^{-5}$ after around $2$ e-folds from the onset of inflation. When the field evolution relaxes to the slow roll attractor, $\epsilon$ approaches a (nearly) constant value of $\epsilon \sim 10^{-10}$, and can be neglected entirely for a period lasting many of e-folds of evolution (Fig. \ref{fig:epsilon_n=3.0115}). The validity of the duality $\eta \rightarrow 3 - \eta$ at the level of the perturbations is clear from the mode equation for the curvature perturbation $\zeta$, 
\begin{equation}
u_k'' + \left[k^2 - \frac{z''}{z}\right] u_k = 0,
\end{equation}
where $u_k = z \zeta_k$, and 
\begin{align}
z = a \frac{\dot{\phi}}{H},
\end{align}
which results in 
\begin{align}
\frac{z''}{z} &= 2 a^2 H^2(1 + \epsilon + \epsilon^2 - \frac{3}{2}\eta + \frac{1}{2}\eta^2 - 2 \epsilon\eta + \frac{1}{2}\xi^2)  \\ \nonumber &=a^2H^2F(\epsilon,\eta,\xi).
\label{eqn:z''/z}
\end{align}
The function $F(\epsilon,\eta,\xi)$ can then be evaluated in the limit of the constant-roll solution (\ref{eq:V_hybrid}), where for $\bar\eta > 0$,
\begin{eqnarray}
&&\epsilon = \bar\eta \tanh{\left(\sqrt{\frac{\bar\eta}{2}} \frac{\phi}{\mpl}\right)}^2\cr
&&\eta = \bar\eta = \mathrm{const.}\cr
&&\xi^2 = \bar\eta^2 \tanh{\left(\sqrt{\frac{\bar\eta}{2}} \frac{\phi}{\mpl}\right)}^2 = \epsilon\bar\eta,
\end{eqnarray}
and for $\bar\eta < 0$,
\begin{eqnarray}
&&\epsilon = - \bar\eta \tan{\left(\sqrt{\frac{-\bar\eta}{2}} \frac{\phi}{\mpl}\right)}^2\cr
&&\eta = \bar\eta = \mathrm{const.}\cr
&&\xi^2 = - \bar\eta^2 \tan{\left(\sqrt{\frac{-\bar\eta}{2}} \frac{\phi}{\mpl}\right)}^2 = \epsilon\bar\eta.
\end{eqnarray}
We then have, for all choices of the potential,
\begin{equation}
F(\epsilon,\bar\eta,\xi) = \left(\eta - 2\right)\left(\eta - 1\right) + \left(\eta^2 - \frac{3 \eta^3}{2}\right) \left(\frac{\phi}{\mpl}\right)^2 + O\left(\frac{\phi}{\mpl}\right)^3 + \cdots,
\end{equation}
which is self-dual under $\eta \rightarrow 3 - \eta$ up to $O\left(\phi/\mpl\right)^2$. This is exactly the ``over the hill'' inflationary evolution considered in Ref. \cite{Tzirakis:2007bf}, where it is shown that self-dual nature of the perturbation solutions holds even in the region where the solution is evolving from the large-$\eta$ solution to the attractor small-$\eta$ solution, with $\eta(t) \neq \mathrm{const.}$, and the power spectrum is a pure power law identical to that given by the slow-roll attractor solution. We note that the background solution is purely classical, and for small enough $\dot\phi$ near the origin, quantum fluctuations will generically dominate, leading to eternal inflation. This case is considered in Refs. \cite{Barenboim:2016mmw,Kinney:2018kew,Jain:2019gsq}.

\section{Phase Space}
\label{sec:PhaseSpace}

In this section we will look at phase space, as done in previous works \cite{Motohashi:2017vdc,Pattison:2018bct}, to assess stability. In phase space parameterized by $(\phi,\dot{\phi})$ the fixed points are the critical points, $V'(\phi) = 0$, of the potential. On a convex potential the critical point is an attractor, where-as on a concave potential it is not. In \cite{Pattison:2018bct} the authors develop a method of parameterizing phase space in terms of $(\phi, \eta(\phi,\dot{\phi}))$ using the equivalent definition of $\eta$, 
\begin{align}
\eta = 3 + \frac{V'}{H\dot{\phi}}.
\label{eq:eta_alt}
\end{align}
A parameterization of this form is more useful for our application since attractor solutions will manifest as the stability of $\eta$. This representation is well defined as long as the field is monotonic, and the mapping $(\phi,\dot{\phi})\rightarrow (\phi,\eta)$ defines an invertible transformation. If this is true, the phase space $(\phi,\eta)$ is equivalent to the phase space defined by $(\phi,\dot{\phi})$ for the purpose of attractor analysis. However, at the critical point, $V'(\phi) =0$, $\eta$ is \textit{no longer} a function of $(\phi,\dot{\phi})$. In particular, at the critical point $\eta$ takes on the value of $\eta = 3$ for all values of $\dot{\phi}$. Therefore, regardless of the field velocity when it evolves though the critical point of the potential, $\eta$ is forced to 3 parametrically, {\it i.e.} there is no other parameter choice. For a generic solution on a given potential the solution will evolve though the critical point and continue evolving on the other side in finite coordinate time. We saw this behavior in our analysis when a perturbation was applied to increased the initial field speed. Phase space analysis in the $\phi,\eta$ plane breaks down at this point because all paths are forced through a single parameter point. Since the paths cross we are no longer able to do phase space analysis. In fact, there exists an infinite set of paths, $(\tilde{\phi},\dot{\phi})$, that cross when $V'(\tilde{\phi})=0$, and regardless of field velocity $\eta = 3$ when the field crosses the critical point of the potential.  

Since our analysis is not about the critical point of the potential, we are able to follow the method developed in Ref. \cite{Pattison:2018bct} to construct phase space diagrams parameterized by $\phi$ and $\eta$, as well as $\phi$ and $\dot{\phi}$. The authors of \cite{Pattison:2018bct} derive an analytical expression linking the potential to the evolution of a field acceleration parameter,
\begin{align}
f = -\frac{\ddot{\phi}}{3 H \dot{\phi}} = \frac{\eta}{3}.
\end{align}
This is equation (3.3) in their work, restated here as 
\begin{align}
\frac{\mathrm{d}f}{\mathrm{d}\phi} = \frac{3}{2\mpl} (f-1)^2(f+1)\left( \frac{V}{V'} \right)\left[\sqrt{1 + \frac{2\mpl^2}{3(f-1)^2} \left(\frac{V'}{V}\right)^2}-\frac{1-f}{1+f} \right] -(1-f) \frac{V''}{V'}.
\label{eq:Analytic_Stability}
\end{align}
It is also possible to link the field speed to the acceleration parameter $f$ though equation (2.6) in \cite{Pattison:2018bct} 
\begin{align}
\dot{\phi}^2 = V \left[ \sqrt{1 + \frac{2 \mpl^2}{3(f-1)^2}\left( \frac{V'}{V}\right)^2}-1 \right].
\label{eq:field_speed}
\end{align}   
From now on we will not work in terms of $f$ but rather $3f = \eta$, since it is more natural in the context of this work. For constant-roll potentials (\ref{eq:Analytic_Stability}) has a fixed point at $3f = \bar{\eta}$. This fixed point will not appear in general potentials, but is unique to constant roll potentials. This fixed point is the advantage of working in the $\phi,\eta$ plane. In the $\phi,\eta$ plane the critical points, which keep $\eta$ fixed, will correspond to a particular path followed by the field. In this way we can analyze the attractor behavior of the constant roll solution. Following from the methodology used in Ref. \cite{Pattison:2018bct}, we will expand around the fixed point
\begin{align}
f\rightarrow \frac{\bar{\eta}}{3} - \delta,
\end{align}
where $\delta$ is a small number, $\delta =(10^{-5})$, in our numerical analysis.

\begin{figure}
\includegraphics[width = 0.8\textwidth]{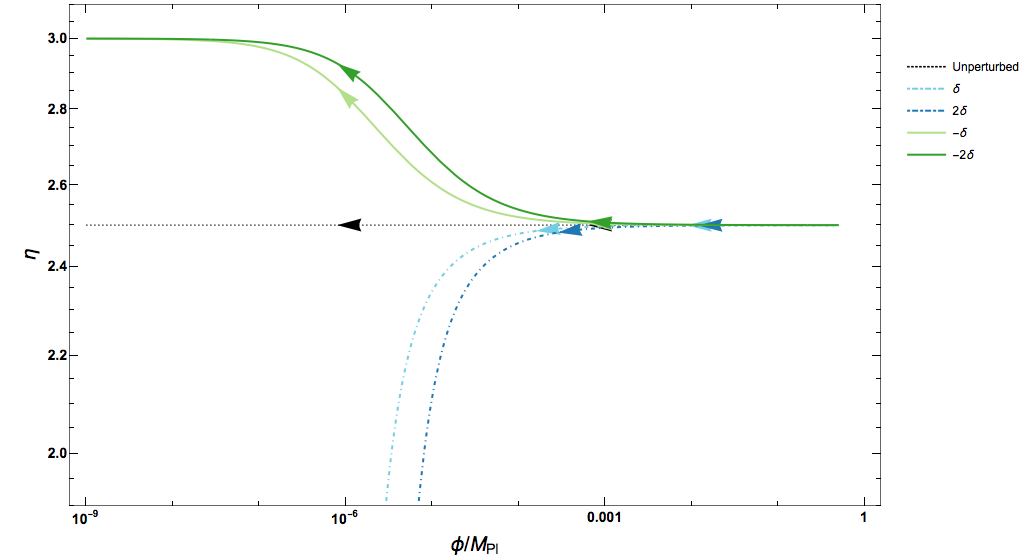}
\caption{Phase space diagram parameterized by $\phi,\eta$ for convex ($\bar{\eta}=2.5$) constant roll potential)}
\label{fig:eta_phi=2_5}
\end{figure}
\begin{figure}
\includegraphics[width = 0.8\textwidth]{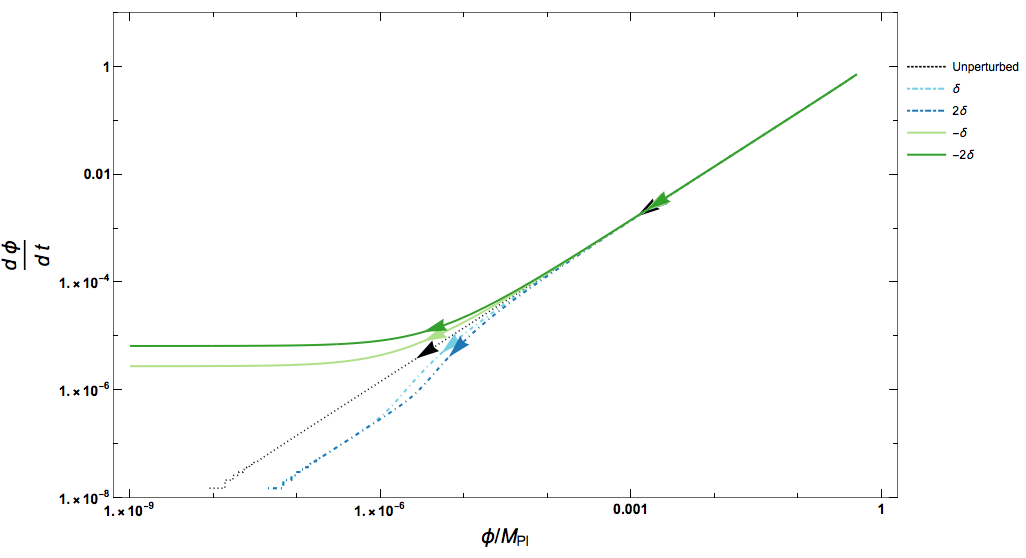}
\caption{Phase space diagram parameterized by $\phi,|\dot{\phi}|$ for convex ($\bar{\eta}=2.5$) constant roll potential)}
\label{fig:speed_phi=2_5}
\end{figure}

\begin{figure}
\includegraphics[width = 0.8\textwidth]{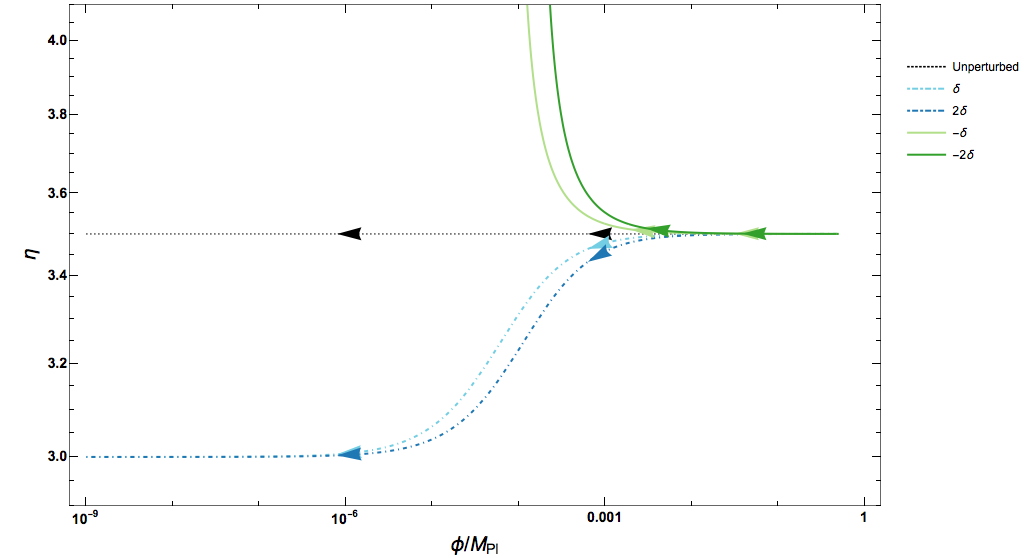}
\caption{Phase space diagram parameterized by $\phi,\eta$ for concave ($\bar{\eta}=3.5$) constant roll potential)}
\label{fig:eta_phi=3_5}
\end{figure}
\begin{figure}
\includegraphics[width = 0.8\textwidth]{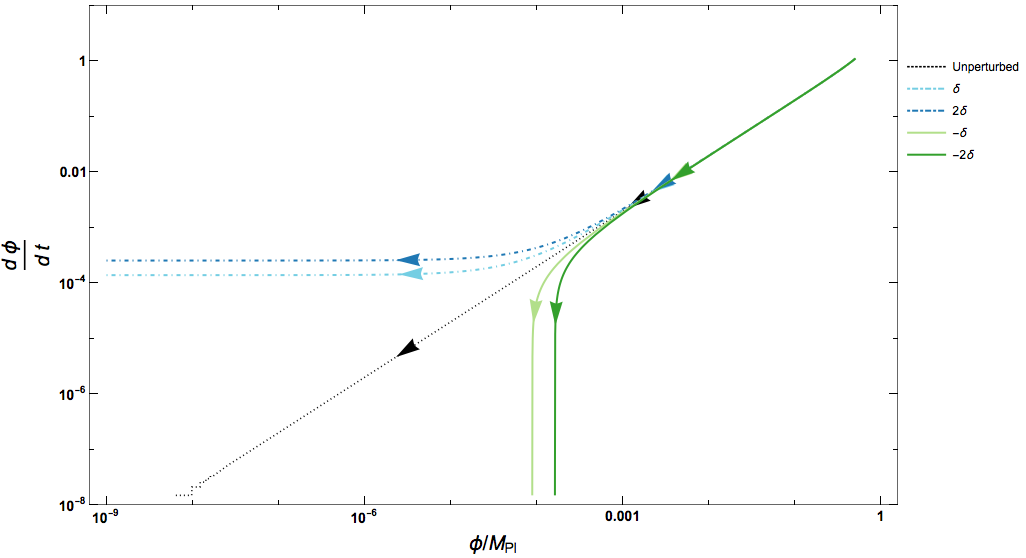}
\caption{Phase space diagram parameterized by $\phi,|\dot{\phi}|$ for concave ($\bar{\eta}=3.5$) constant roll potential)}
\label{fig:speed_phi=3_5}
\end{figure}

For the convex ($\bar{\eta} = 2.5$) potential case we considered in Sec. \ref{subsec:Hybrid} we analyze the $(\phi,\eta)$ plane and accompanying $(\phi,\dot{\phi})$ plane though Eqs. (\ref{eq:Analytic_Stability}) and (\ref{eq:field_speed}). Our analysis is done numerically and plotted in (\ref{fig:eta_phi=2_5}) and (\ref{fig:speed_phi=2_5}). In Fig. (\ref{fig:eta_phi=2_5}) one clearly sees that as $\phi \rightarrow 0$, $\eta$ evolves away from the fixed $\eta = \bar{\eta}$ solution and either to slow roll (dashed) or to $\eta = 3$ (solid). As explained in the earlier section, the dashed lines correspond to a decreased field speed such that the field will come to rest before passing the critical point of the potential in slow roll $|\eta|<1$. Where as, the solid lines correspond to the field rolling past the potential's fixed point, eventually it will turn around and roll back to the minimum in slow roll. This behavior is clearly seen in Fig. (\ref{fig:speed_phi=2_5}) where the fields corresponding to the solid lines have finite speed about $\phi = 0$, and the fields corresponding to the dashed lines have speeds asymptoticly approaching zero.

The same analysis is done again for the concave potential ($\bar{\eta} = 3.5$) discussed in Sec. \ref{subsec:Hybrid}. The numerical results are presented in Figs. (\ref{fig:eta_phi=3_5}) and (\ref{fig:speed_phi=3_5}). Again one sees that as $\phi \rightarrow 0$, $\eta$ will evolve away from the fixed $\eta = \bar{\eta}$ solution either into slow roll (solid) or to $\eta = 3$ (dashed). Since the potential is concave the solid lines represent a field that comes to rest before the maximum of the potential and rolls back down. One can see in Fig. (\ref{fig:speed_phi=3_5}) that the fields corresponding to the solid lines have speeds that quickly approach zero. The other option, the dashed lines, correspond to fields that roll over the top of the potential, and as expected they keep finite field speed as they approach it.

In each case we see that as the field evolves toward the origin, $|\delta|$ will grow causing $\eta$ to diverge away from the constant roll value $\bar{\eta}$. From these phase diagrams it may seem that an attractor exists at $\eta = 3$, however this is due to the fact that when the field is at the critical point $\eta = 3$ regardless of the field velocity. The limit $\eta = 3$ is not the late time attractor of these fields due to the fact they will roll past the critical point and continue to evolve. This is verified by looking at the field speed as $\phi \rightarrow 0$ as is done in Figs. (\ref{fig:speed_phi=2_5}) and (\ref{fig:speed_phi=3_5}) where we look at the phase space parametrized by $(\phi,|\dot{\phi}|)$. The field is approaching the critical point, $V'=0,\eta = 3$, with finite speed and will continue past it. At the point that the field passes through the critical point, $V'=0$, $\eta$ is forced to 3 by definition (\ref{eq:eta_alt}), therefore while on the phase diagram $\eta = 3$ appears to be an attractor, it is not one in the normal sense.  This phase space analysis confirm the results of our analysis done in the $t,\eta$ plane that the constant $\eta$ solutions are not attractors.   

\section{Conclusion}
\label{sec:Conclusion}

In this paper, we clarify the large-$\eta$ attractor behavior of constant roll inflation and the duality relation $\eta \rightarrow 3 - \eta$. We show that for both $0<\epsilon<\eta<3$ convex potentials and $\eta>3$ concave potentials, perturbations to the analytic constant roll solution will result in $\eta$ evolving to the smaller of $\eta$, and $\tilde{\eta}=3-\eta$. We show that for the phenomenologically viable $\eta \approx 3.0115$ case considered by Gao, {\it et al.}, \cite{Gao:2019sbz}, the background solution is unstable, and we introduce finely tuned initial conditions such that the field evolves keeping $\eta = \bar{\eta} = $ constant. Generically, small perturbations to the analytic constant roll solution will cause the dynamic parameter $\eta$ to evolve away from the larger, unstable, value of $ \left\lbrace  \bar{\eta},\ 3- \bar{\eta} \right \rbrace$, to the smaller value, which is the stable solution.  This behavior can be easily understood by analyzing the field EOM as it evolves toward the linear region near the extremum of the potential. Since $H$ approaches a constant exponentially quickly by the time the field is in the small-field limit, the field EOM becomes a linear second order differential equation, with two linearly independent solutions defining the solution space. Perturbations to the large-$\bar{\eta}$ analytic constant roll solution correspond to excitations of the other linearly independent term (\ref{eq:smallFieldSolution}) with $\tilde{\eta} = 3- \bar{\eta}$,  which will dominate the field evolution.

Since the large-$\eta$ analytic constant roll solutions are not stable under perturbations, they do not constitute late-time attractors. We can in some sense consider the large-$\eta$ solutions as early time transients, since the field may evolve with large-$\eta$ for a few e-folds, until it enters the linear regime and the evolution is dominated by slow roll.

\begin{acknowledgments}
WHK is supported by the Vetenskapsr\r{a}det (Swedish Research Council) through contract No. 638-2013-8993 and the Oskar Klein Centre for Cosmoparticle Physics, and by the U.S. National Science Foundation under grant NSF-PHY-1719690. 
\end{acknowledgments}

\bibliographystyle{apsrev4-1}
\bibliography{paper}

\begin{thebibliography}{65}%
\makeatletter
\providecommand \@ifxundefined [1]{%
 \@ifx{#1\undefined}
}%
\providecommand \@ifnum [1]{%
 \ifnum #1\expandafter \@firstoftwo
 \else \expandafter \@secondoftwo
 \fi
}%
\providecommand \@ifx [1]{%
 \ifx #1\expandafter \@firstoftwo
 \else \expandafter \@secondoftwo
 \fi
}%
\providecommand \natexlab [1]{#1}%
\providecommand \enquote  [1]{``#1''}%
\providecommand \bibnamefont  [1]{#1}%
\providecommand \bibfnamefont [1]{#1}%
\providecommand \citenamefont [1]{#1}%
\providecommand \href@noop [0]{\@secondoftwo}%
\providecommand \href [0]{\begingroup \@sanitize@url \@href}%
\providecommand \@href[1]{\@@startlink{#1}\@@href}%
\providecommand \@@href[1]{\endgroup#1\@@endlink}%
\providecommand \@sanitize@url [0]{\catcode `\\12\catcode `\$12\catcode
  `\&12\catcode `\#12\catcode `\^12\catcode `\_12\catcode `\%12\relax}%
\providecommand \@@startlink[1]{}%
\providecommand \@@endlink[0]{}%
\providecommand \url  [0]{\begingroup\@sanitize@url \@url }%
\providecommand \@url [1]{\endgroup\@href {#1}{\urlprefix }}%
\providecommand \urlprefix  [0]{URL }%
\providecommand \Eprint [0]{\href }%
\providecommand \doibase [0]{http://dx.doi.org/}%
\providecommand \selectlanguage [0]{\@gobble}%
\providecommand \bibinfo  [0]{\@secondoftwo}%
\providecommand \bibfield  [0]{\@secondoftwo}%
\providecommand \translation [1]{[#1]}%
\providecommand \BibitemOpen [0]{}%
\providecommand \bibitemStop [0]{}%
\providecommand \bibitemNoStop [0]{.\EOS\space}%
\providecommand \EOS [0]{\spacefactor3000\relax}%
\providecommand \BibitemShut  [1]{\csname bibitem#1\endcsname}%
\let\auto@bib@innerbib\@empty
\bibitem [{\citenamefont {Motohashi}\ \emph {et~al.}(2015)\citenamefont
  {Motohashi}, \citenamefont {Starobinsky},\ and\ \citenamefont
  {Yokoyama}}]{Motohashi:2014ppa}%
  \BibitemOpen
  \bibfield  {author} {\bibinfo {author} {\bibfnamefont {H.}~\bibnamefont
  {Motohashi}}, \bibinfo {author} {\bibfnamefont {A.~A.}\ \bibnamefont
  {Starobinsky}}, \ and\ \bibinfo {author} {\bibfnamefont {J.}~\bibnamefont
  {Yokoyama}},\ }\href {\doibase 10.1088/1475-7516/2015/09/018} {\bibfield
  {journal} {\bibinfo  {journal} {JCAP}\ }\textbf {\bibinfo {volume} {1509}},\
  \bibinfo {pages} {018} (\bibinfo {year} {2015})},\ \Eprint
  {http://arxiv.org/abs/1411.5021} {arXiv:1411.5021 [astro-ph.CO]} \BibitemShut
  {NoStop}%
\bibitem [{\citenamefont {Kinney}(1997)}]{Kinney:1997ne}%
  \BibitemOpen
  \bibfield  {author} {\bibinfo {author} {\bibfnamefont {W.~H.}\ \bibnamefont
  {Kinney}},\ }\href {\doibase 10.1103/PhysRevD.56.2002} {\bibfield  {journal}
  {\bibinfo  {journal} {Phys. Rev.}\ }\textbf {\bibinfo {volume} {D56}},\
  \bibinfo {pages} {2002} (\bibinfo {year} {1997})},\ \Eprint
  {http://arxiv.org/abs/hep-ph/9702427} {arXiv:hep-ph/9702427 [hep-ph]}
  \BibitemShut {NoStop}%
\bibitem [{\citenamefont {Tsamis}\ and\ \citenamefont
  {Woodard}(2004)}]{Tsamis:2003px}%
  \BibitemOpen
  \bibfield  {author} {\bibinfo {author} {\bibfnamefont {N.~C.}\ \bibnamefont
  {Tsamis}}\ and\ \bibinfo {author} {\bibfnamefont {R.~P.}\ \bibnamefont
  {Woodard}},\ }\href {\doibase 10.1103/PhysRevD.69.084005} {\bibfield
  {journal} {\bibinfo  {journal} {Phys. Rev.}\ }\textbf {\bibinfo {volume}
  {D69}},\ \bibinfo {pages} {084005} (\bibinfo {year} {2004})},\ \Eprint
  {http://arxiv.org/abs/astro-ph/0307463} {arXiv:astro-ph/0307463 [astro-ph]}
  \BibitemShut {NoStop}%
\bibitem [{\citenamefont {Kinney}(2005)}]{Kinney:2005vj}%
  \BibitemOpen
  \bibfield  {author} {\bibinfo {author} {\bibfnamefont {W.~H.}\ \bibnamefont
  {Kinney}},\ }\href {\doibase 10.1103/PhysRevD.72.023515} {\bibfield
  {journal} {\bibinfo  {journal} {Phys. Rev.}\ }\textbf {\bibinfo {volume}
  {D72}},\ \bibinfo {pages} {023515} (\bibinfo {year} {2005})},\ \Eprint
  {http://arxiv.org/abs/gr-qc/0503017} {arXiv:gr-qc/0503017 [gr-qc]}
  \BibitemShut {NoStop}%
\bibitem [{\citenamefont {Martin}\ \emph {et~al.}(2013)\citenamefont {Martin},
  \citenamefont {Motohashi},\ and\ \citenamefont {Suyama}}]{Martin:2012pe}%
  \BibitemOpen
  \bibfield  {author} {\bibinfo {author} {\bibfnamefont {J.}~\bibnamefont
  {Martin}}, \bibinfo {author} {\bibfnamefont {H.}~\bibnamefont {Motohashi}}, \
  and\ \bibinfo {author} {\bibfnamefont {T.}~\bibnamefont {Suyama}},\ }\href
  {\doibase 10.1103/PhysRevD.87.023514} {\bibfield  {journal} {\bibinfo
  {journal} {Phys. Rev.}\ }\textbf {\bibinfo {volume} {D87}},\ \bibinfo {pages}
  {023514} (\bibinfo {year} {2013})},\ \Eprint {http://arxiv.org/abs/1211.0083}
  {arXiv:1211.0083 [astro-ph.CO]} \BibitemShut {NoStop}%
\bibitem [{\citenamefont {Inoue}\ and\ \citenamefont
  {Yokoyama}(2002)}]{Inoue:2001zt}%
  \BibitemOpen
  \bibfield  {author} {\bibinfo {author} {\bibfnamefont {S.}~\bibnamefont
  {Inoue}}\ and\ \bibinfo {author} {\bibfnamefont {J.}~\bibnamefont
  {Yokoyama}},\ }\href {\doibase 10.1016/S0370-2693(01)01369-7} {\bibfield
  {journal} {\bibinfo  {journal} {Phys. Lett.}\ }\textbf {\bibinfo {volume}
  {B524}},\ \bibinfo {pages} {15} (\bibinfo {year} {2002})},\ \Eprint
  {http://arxiv.org/abs/hep-ph/0104083} {arXiv:hep-ph/0104083 [hep-ph]}
  \BibitemShut {NoStop}%
\bibitem [{\citenamefont {Namjoo}\ \emph {et~al.}(2013)\citenamefont {Namjoo},
  \citenamefont {Firouzjahi},\ and\ \citenamefont {Sasaki}}]{Namjoo:2012aa}%
  \BibitemOpen
  \bibfield  {author} {\bibinfo {author} {\bibfnamefont {M.~H.}\ \bibnamefont
  {Namjoo}}, \bibinfo {author} {\bibfnamefont {H.}~\bibnamefont {Firouzjahi}},
  \ and\ \bibinfo {author} {\bibfnamefont {M.}~\bibnamefont {Sasaki}},\ }\href
  {\doibase 10.1209/0295-5075/101/39001} {\bibfield  {journal} {\bibinfo
  {journal} {EPL}\ }\textbf {\bibinfo {volume} {101}},\ \bibinfo {pages}
  {39001} (\bibinfo {year} {2013})},\ \Eprint {http://arxiv.org/abs/1210.3692}
  {arXiv:1210.3692 [astro-ph.CO]} \BibitemShut {NoStop}%
\bibitem [{\citenamefont {Huang}\ and\ \citenamefont
  {Wang}(2013)}]{Huang:2013lda}%
  \BibitemOpen
  \bibfield  {author} {\bibinfo {author} {\bibfnamefont {Q.-G.}\ \bibnamefont
  {Huang}}\ and\ \bibinfo {author} {\bibfnamefont {Y.}~\bibnamefont {Wang}},\
  }\href {\doibase 10.1088/1475-7516/2013/06/035} {\bibfield  {journal}
  {\bibinfo  {journal} {JCAP}\ }\textbf {\bibinfo {volume} {1306}},\ \bibinfo
  {pages} {035} (\bibinfo {year} {2013})},\ \Eprint
  {http://arxiv.org/abs/1303.4526} {arXiv:1303.4526 [hep-th]} \BibitemShut
  {NoStop}%
\bibitem [{\citenamefont {Mooij}\ and\ \citenamefont
  {Palma}(2015)}]{Mooij:2015yka}%
  \BibitemOpen
  \bibfield  {author} {\bibinfo {author} {\bibfnamefont {S.}~\bibnamefont
  {Mooij}}\ and\ \bibinfo {author} {\bibfnamefont {G.~A.}\ \bibnamefont
  {Palma}},\ }\href {\doibase 10.1088/1475-7516/2015/11/025} {\bibfield
  {journal} {\bibinfo  {journal} {JCAP}\ }\textbf {\bibinfo {volume} {1511}},\
  \bibinfo {pages} {025} (\bibinfo {year} {2015})},\ \Eprint
  {http://arxiv.org/abs/1502.03458} {arXiv:1502.03458 [astro-ph.CO]}
  \BibitemShut {NoStop}%
\bibitem [{\citenamefont {Cicciarella}\ \emph {et~al.}(2018)\citenamefont
  {Cicciarella}, \citenamefont {Mabillard},\ and\ \citenamefont
  {Pieroni}}]{Cicciarella:2017nls}%
  \BibitemOpen
  \bibfield  {author} {\bibinfo {author} {\bibfnamefont {F.}~\bibnamefont
  {Cicciarella}}, \bibinfo {author} {\bibfnamefont {J.}~\bibnamefont
  {Mabillard}}, \ and\ \bibinfo {author} {\bibfnamefont {M.}~\bibnamefont
  {Pieroni}},\ }\href {\doibase 10.1088/1475-7516/2018/01/024} {\bibfield
  {journal} {\bibinfo  {journal} {JCAP}\ }\textbf {\bibinfo {volume} {1801}},\
  \bibinfo {pages} {024} (\bibinfo {year} {2018})},\ \Eprint
  {http://arxiv.org/abs/1709.03527} {arXiv:1709.03527 [astro-ph.CO]}
  \BibitemShut {NoStop}%
\bibitem [{\citenamefont {Akhshik}\ \emph {et~al.}(2015)\citenamefont
  {Akhshik}, \citenamefont {Firouzjahi},\ and\ \citenamefont
  {Jazayeri}}]{Akhshik:2015nfa}%
  \BibitemOpen
  \bibfield  {author} {\bibinfo {author} {\bibfnamefont {M.}~\bibnamefont
  {Akhshik}}, \bibinfo {author} {\bibfnamefont {H.}~\bibnamefont {Firouzjahi}},
  \ and\ \bibinfo {author} {\bibfnamefont {S.}~\bibnamefont {Jazayeri}},\
  }\href {\doibase 10.1088/1475-7516/2015/07/048} {\bibfield  {journal}
  {\bibinfo  {journal} {JCAP}\ }\textbf {\bibinfo {volume} {1507}},\ \bibinfo
  {pages} {048} (\bibinfo {year} {2015})},\ \Eprint
  {http://arxiv.org/abs/1501.01099} {arXiv:1501.01099 [hep-th]} \BibitemShut
  {NoStop}%
\bibitem [{\citenamefont {Scacco}\ and\ \citenamefont
  {Albrecht}(2015)}]{Scacco:2015spa}%
  \BibitemOpen
  \bibfield  {author} {\bibinfo {author} {\bibfnamefont {A.}~\bibnamefont
  {Scacco}}\ and\ \bibinfo {author} {\bibfnamefont {A.}~\bibnamefont
  {Albrecht}},\ }\href {\doibase 10.1103/PhysRevD.92.083506} {\bibfield
  {journal} {\bibinfo  {journal} {Phys. Rev.}\ }\textbf {\bibinfo {volume}
  {D92}},\ \bibinfo {pages} {083506} (\bibinfo {year} {2015})},\ \Eprint
  {http://arxiv.org/abs/1503.04872} {arXiv:1503.04872 [astro-ph.CO]}
  \BibitemShut {NoStop}%
\bibitem [{\citenamefont {Barenboim}\ \emph {et~al.}(2016)\citenamefont
  {Barenboim}, \citenamefont {Park},\ and\ \citenamefont
  {Kinney}}]{Barenboim:2016mmw}%
  \BibitemOpen
  \bibfield  {author} {\bibinfo {author} {\bibfnamefont {G.}~\bibnamefont
  {Barenboim}}, \bibinfo {author} {\bibfnamefont {W.-I.}\ \bibnamefont {Park}},
  \ and\ \bibinfo {author} {\bibfnamefont {W.~H.}\ \bibnamefont {Kinney}},\
  }\href {\doibase 10.1088/1475-7516/2016/05/030} {\bibfield  {journal}
  {\bibinfo  {journal} {JCAP}\ }\textbf {\bibinfo {volume} {1605}},\ \bibinfo
  {pages} {030} (\bibinfo {year} {2016})},\ \Eprint
  {http://arxiv.org/abs/1601.08140} {arXiv:1601.08140 [astro-ph.CO]}
  \BibitemShut {NoStop}%
\bibitem [{\citenamefont {Cai}\ \emph {et~al.}(2016)\citenamefont {Cai},
  \citenamefont {Gong}, \citenamefont {Wang},\ and\ \citenamefont
  {Wang}}]{Cai:2016ngx}%
  \BibitemOpen
  \bibfield  {author} {\bibinfo {author} {\bibfnamefont {Y.-F.}\ \bibnamefont
  {Cai}}, \bibinfo {author} {\bibfnamefont {J.-O.}\ \bibnamefont {Gong}},
  \bibinfo {author} {\bibfnamefont {D.-G.}\ \bibnamefont {Wang}}, \ and\
  \bibinfo {author} {\bibfnamefont {Z.}~\bibnamefont {Wang}},\ }\href {\doibase
  10.1088/1475-7516/2016/10/017} {\bibfield  {journal} {\bibinfo  {journal}
  {JCAP}\ }\textbf {\bibinfo {volume} {1610}},\ \bibinfo {pages} {017}
  (\bibinfo {year} {2016})},\ \Eprint {http://arxiv.org/abs/1607.07872}
  {arXiv:1607.07872 [astro-ph.CO]} \BibitemShut {NoStop}%
\bibitem [{\citenamefont {Odintsov}\ and\ \citenamefont
  {Oikonomou}(2017{\natexlab{a}})}]{Odintsov:2017yud}%
  \BibitemOpen
  \bibfield  {author} {\bibinfo {author} {\bibfnamefont {S.~D.}\ \bibnamefont
  {Odintsov}}\ and\ \bibinfo {author} {\bibfnamefont {V.~K.}\ \bibnamefont
  {Oikonomou}},\ }\href {\doibase 10.1088/1475-7516/2017/04/041} {\bibfield
  {journal} {\bibinfo  {journal} {JCAP}\ }\textbf {\bibinfo {volume} {1704}},\
  \bibinfo {pages} {041} (\bibinfo {year} {2017}{\natexlab{a}})},\ \Eprint
  {http://arxiv.org/abs/1703.02853} {arXiv:1703.02853 [gr-qc]} \BibitemShut
  {NoStop}%
\bibitem [{\citenamefont {Grain}\ and\ \citenamefont
  {Vennin}(2017)}]{Grain:2017dqa}%
  \BibitemOpen
  \bibfield  {author} {\bibinfo {author} {\bibfnamefont {J.}~\bibnamefont
  {Grain}}\ and\ \bibinfo {author} {\bibfnamefont {V.}~\bibnamefont {Vennin}},\
  }\href {\doibase 10.1088/1475-7516/2017/05/045} {\bibfield  {journal}
  {\bibinfo  {journal} {JCAP}\ }\textbf {\bibinfo {volume} {1705}},\ \bibinfo
  {pages} {045} (\bibinfo {year} {2017})},\ \Eprint
  {http://arxiv.org/abs/1703.00447} {arXiv:1703.00447 [gr-qc]} \BibitemShut
  {NoStop}%
\bibitem [{\citenamefont {Odintsov}\ and\ \citenamefont
  {Oikonomou}(2017{\natexlab{b}})}]{Odintsov:2017qpp}%
  \BibitemOpen
  \bibfield  {author} {\bibinfo {author} {\bibfnamefont {S.~D.}\ \bibnamefont
  {Odintsov}}\ and\ \bibinfo {author} {\bibfnamefont {V.~K.}\ \bibnamefont
  {Oikonomou}},\ }\href {\doibase 10.1103/PhysRevD.96.024029} {\bibfield
  {journal} {\bibinfo  {journal} {Phys. Rev.}\ }\textbf {\bibinfo {volume}
  {D96}},\ \bibinfo {pages} {024029} (\bibinfo {year} {2017}{\natexlab{b}})},\
  \Eprint {http://arxiv.org/abs/1704.02931} {arXiv:1704.02931 [gr-qc]}
  \BibitemShut {NoStop}%
\bibitem [{\citenamefont {Bravo}\ \emph
  {et~al.}(2018{\natexlab{a}})\citenamefont {Bravo}, \citenamefont {Mooij},
  \citenamefont {Palma},\ and\ \citenamefont {Pradenas}}]{Bravo:2017wyw}%
  \BibitemOpen
  \bibfield  {author} {\bibinfo {author} {\bibfnamefont {R.}~\bibnamefont
  {Bravo}}, \bibinfo {author} {\bibfnamefont {S.}~\bibnamefont {Mooij}},
  \bibinfo {author} {\bibfnamefont {G.~A.}\ \bibnamefont {Palma}}, \ and\
  \bibinfo {author} {\bibfnamefont {B.}~\bibnamefont {Pradenas}},\ }\href
  {\doibase 10.1088/1475-7516/2018/05/024} {\bibfield  {journal} {\bibinfo
  {journal} {JCAP}\ }\textbf {\bibinfo {volume} {1805}},\ \bibinfo {pages}
  {024} (\bibinfo {year} {2018}{\natexlab{a}})},\ \Eprint
  {http://arxiv.org/abs/1711.02680} {arXiv:1711.02680 [astro-ph.CO]}
  \BibitemShut {NoStop}%
\bibitem [{\citenamefont {Bravo}\ \emph
  {et~al.}(2018{\natexlab{b}})\citenamefont {Bravo}, \citenamefont {Mooij},
  \citenamefont {Palma},\ and\ \citenamefont {Pradenas}}]{Bravo:2017gct}%
  \BibitemOpen
  \bibfield  {author} {\bibinfo {author} {\bibfnamefont {R.}~\bibnamefont
  {Bravo}}, \bibinfo {author} {\bibfnamefont {S.}~\bibnamefont {Mooij}},
  \bibinfo {author} {\bibfnamefont {G.~A.}\ \bibnamefont {Palma}}, \ and\
  \bibinfo {author} {\bibfnamefont {B.}~\bibnamefont {Pradenas}},\ }\href
  {\doibase 10.1088/1475-7516/2018/05/025} {\bibfield  {journal} {\bibinfo
  {journal} {JCAP}\ }\textbf {\bibinfo {volume} {1805}},\ \bibinfo {pages}
  {025} (\bibinfo {year} {2018}{\natexlab{b}})},\ \Eprint
  {http://arxiv.org/abs/1711.05290} {arXiv:1711.05290 [astro-ph.CO]}
  \BibitemShut {NoStop}%
\bibitem [{\citenamefont {Dimopoulos}(2017)}]{Dimopoulos:2017ged}%
  \BibitemOpen
  \bibfield  {author} {\bibinfo {author} {\bibfnamefont {K.}~\bibnamefont
  {Dimopoulos}},\ }\href {\doibase 10.1016/j.physletb.2017.10.066} {\bibfield
  {journal} {\bibinfo  {journal} {Phys. Lett.}\ }\textbf {\bibinfo {volume}
  {B775}},\ \bibinfo {pages} {262} (\bibinfo {year} {2017})},\ \Eprint
  {http://arxiv.org/abs/1707.05644} {arXiv:1707.05644 [hep-ph]} \BibitemShut
  {NoStop}%
\bibitem [{\citenamefont {Nojiri}\ \emph {et~al.}(2017)\citenamefont {Nojiri},
  \citenamefont {Odintsov},\ and\ \citenamefont {Oikonomou}}]{Nojiri:2017qvx}%
  \BibitemOpen
  \bibfield  {author} {\bibinfo {author} {\bibfnamefont {S.}~\bibnamefont
  {Nojiri}}, \bibinfo {author} {\bibfnamefont {S.~D.}\ \bibnamefont
  {Odintsov}}, \ and\ \bibinfo {author} {\bibfnamefont {V.~K.}\ \bibnamefont
  {Oikonomou}},\ }\href {\doibase 10.1088/1361-6382/aa92a4} {\bibfield
  {journal} {\bibinfo  {journal} {Class. Quant. Grav.}\ }\textbf {\bibinfo
  {volume} {34}},\ \bibinfo {pages} {245012} (\bibinfo {year} {2017})},\
  \Eprint {http://arxiv.org/abs/1704.05945} {arXiv:1704.05945 [gr-qc]}
  \BibitemShut {NoStop}%
\bibitem [{\citenamefont {Motohashi}\ and\ \citenamefont
  {Starobinsky}(2017)}]{Motohashi:2017vdc}%
  \BibitemOpen
  \bibfield  {author} {\bibinfo {author} {\bibfnamefont {H.}~\bibnamefont
  {Motohashi}}\ and\ \bibinfo {author} {\bibfnamefont {A.~A.}\ \bibnamefont
  {Starobinsky}},\ }\href {\doibase 10.1140/epjc/s10052-017-5109-x} {\bibfield
  {journal} {\bibinfo  {journal} {Eur. Phys. J.}\ }\textbf {\bibinfo {volume}
  {C77}},\ \bibinfo {pages} {538} (\bibinfo {year} {2017})},\ \Eprint
  {http://arxiv.org/abs/1704.08188} {arXiv:1704.08188 [astro-ph.CO]}
  \BibitemShut {NoStop}%
\bibitem [{\citenamefont {Odintsov}\ \emph {et~al.}(2017)\citenamefont
  {Odintsov}, \citenamefont {Oikonomou},\ and\ \citenamefont
  {Sebastiani}}]{Odintsov:2017hbk}%
  \BibitemOpen
  \bibfield  {author} {\bibinfo {author} {\bibfnamefont {S.~D.}\ \bibnamefont
  {Odintsov}}, \bibinfo {author} {\bibfnamefont {V.~K.}\ \bibnamefont
  {Oikonomou}}, \ and\ \bibinfo {author} {\bibfnamefont {L.}~\bibnamefont
  {Sebastiani}},\ }\href {\doibase 10.1016/j.nuclphysb.2017.08.018} {\bibfield
  {journal} {\bibinfo  {journal} {Nucl. Phys.}\ }\textbf {\bibinfo {volume}
  {B923}},\ \bibinfo {pages} {608} (\bibinfo {year} {2017})},\ \Eprint
  {http://arxiv.org/abs/1708.08346} {arXiv:1708.08346 [gr-qc]} \BibitemShut
  {NoStop}%
\bibitem [{\citenamefont {Oikonomou}(2017)}]{Oikonomou:2017xik}%
  \BibitemOpen
  \bibfield  {author} {\bibinfo {author} {\bibfnamefont {V.~K.}\ \bibnamefont
  {Oikonomou}},\ }\href {\doibase 10.1142/S0218271818500098} {\bibfield
  {journal} {\bibinfo  {journal} {Int. J. Mod. Phys.}\ }\textbf {\bibinfo
  {volume} {D27}},\ \bibinfo {pages} {1850009} (\bibinfo {year} {2017})},\
  \Eprint {http://arxiv.org/abs/1709.02986} {arXiv:1709.02986 [gr-qc]}
  \BibitemShut {NoStop}%
\bibitem [{\citenamefont {Awad}\ \emph {et~al.}(2018)\citenamefont {Awad},
  \citenamefont {El~Hanafy}, \citenamefont {Nashed}, \citenamefont {Odintsov},\
  and\ \citenamefont {Oikonomou}}]{Awad:2017ign}%
  \BibitemOpen
  \bibfield  {author} {\bibinfo {author} {\bibfnamefont {A.}~\bibnamefont
  {Awad}}, \bibinfo {author} {\bibfnamefont {W.}~\bibnamefont {El~Hanafy}},
  \bibinfo {author} {\bibfnamefont {G.~G.~L.}\ \bibnamefont {Nashed}}, \bibinfo
  {author} {\bibfnamefont {S.~D.}\ \bibnamefont {Odintsov}}, \ and\ \bibinfo
  {author} {\bibfnamefont {V.~K.}\ \bibnamefont {Oikonomou}},\ }\href {\doibase
  10.1088/1475-7516/2018/07/026} {\bibfield  {journal} {\bibinfo  {journal}
  {JCAP}\ }\textbf {\bibinfo {volume} {1807}},\ \bibinfo {pages} {026}
  (\bibinfo {year} {2018})},\ \Eprint {http://arxiv.org/abs/1710.00682}
  {arXiv:1710.00682 [gr-qc]} \BibitemShut {NoStop}%
\bibitem [{\citenamefont {Anguelova}\ \emph {et~al.}(2018)\citenamefont
  {Anguelova}, \citenamefont {Suranyi},\ and\ \citenamefont
  {Wijewardhana}}]{Anguelova:2017djf}%
  \BibitemOpen
  \bibfield  {author} {\bibinfo {author} {\bibfnamefont {L.}~\bibnamefont
  {Anguelova}}, \bibinfo {author} {\bibfnamefont {P.}~\bibnamefont {Suranyi}},
  \ and\ \bibinfo {author} {\bibfnamefont {L.~C.~R.}\ \bibnamefont
  {Wijewardhana}},\ }\href {\doibase 10.1088/1475-7516/2018/02/004} {\bibfield
  {journal} {\bibinfo  {journal} {JCAP}\ }\textbf {\bibinfo {volume} {1802}},\
  \bibinfo {pages} {004} (\bibinfo {year} {2018})},\ \Eprint
  {http://arxiv.org/abs/1710.06989} {arXiv:1710.06989 [hep-th]} \BibitemShut
  {NoStop}%
\bibitem [{\citenamefont {Salvio}(2018)}]{Salvio:2017oyf}%
  \BibitemOpen
  \bibfield  {author} {\bibinfo {author} {\bibfnamefont {A.}~\bibnamefont
  {Salvio}},\ }\href {\doibase 10.1016/j.physletb.2018.03.009} {\bibfield
  {journal} {\bibinfo  {journal} {Phys. Lett.}\ }\textbf {\bibinfo {volume}
  {B780}},\ \bibinfo {pages} {111} (\bibinfo {year} {2018})},\ \Eprint
  {http://arxiv.org/abs/1712.04477} {arXiv:1712.04477 [hep-ph]} \BibitemShut
  {NoStop}%
\bibitem [{\citenamefont {Yi}\ and\ \citenamefont {Gong}(2018)}]{Yi:2017mxs}%
  \BibitemOpen
  \bibfield  {author} {\bibinfo {author} {\bibfnamefont {Z.}~\bibnamefont
  {Yi}}\ and\ \bibinfo {author} {\bibfnamefont {Y.}~\bibnamefont {Gong}},\
  }\href {\doibase 10.1088/1475-7516/2018/03/052} {\bibfield  {journal}
  {\bibinfo  {journal} {JCAP}\ }\textbf {\bibinfo {volume} {1803}},\ \bibinfo
  {pages} {052} (\bibinfo {year} {2018})},\ \Eprint
  {http://arxiv.org/abs/1712.07478} {arXiv:1712.07478 [gr-qc]} \BibitemShut
  {NoStop}%
\bibitem [{\citenamefont {Cai}\ \emph {et~al.}(2018)\citenamefont {Cai},
  \citenamefont {Chen}, \citenamefont {Namjoo}, \citenamefont {Sasaki},
  \citenamefont {Wang},\ and\ \citenamefont {Wang}}]{Cai:2017bxr}%
  \BibitemOpen
  \bibfield  {author} {\bibinfo {author} {\bibfnamefont {Y.-F.}\ \bibnamefont
  {Cai}}, \bibinfo {author} {\bibfnamefont {X.}~\bibnamefont {Chen}}, \bibinfo
  {author} {\bibfnamefont {M.~H.}\ \bibnamefont {Namjoo}}, \bibinfo {author}
  {\bibfnamefont {M.}~\bibnamefont {Sasaki}}, \bibinfo {author} {\bibfnamefont
  {D.-G.}\ \bibnamefont {Wang}}, \ and\ \bibinfo {author} {\bibfnamefont
  {Z.}~\bibnamefont {Wang}},\ }\href {\doibase 10.1088/1475-7516/2018/05/012}
  {\bibfield  {journal} {\bibinfo  {journal} {JCAP}\ }\textbf {\bibinfo
  {volume} {1805}},\ \bibinfo {pages} {012} (\bibinfo {year} {2018})},\ \Eprint
  {http://arxiv.org/abs/1712.09998} {arXiv:1712.09998 [astro-ph.CO]}
  \BibitemShut {NoStop}%
\bibitem [{\citenamefont {Mohammadi}\ \emph
  {et~al.}(2018{\natexlab{a}})\citenamefont {Mohammadi}, \citenamefont
  {Saaidi},\ and\ \citenamefont {Golanbari}}]{Mohammadi:2018oku}%
  \BibitemOpen
  \bibfield  {author} {\bibinfo {author} {\bibfnamefont {A.}~\bibnamefont
  {Mohammadi}}, \bibinfo {author} {\bibfnamefont {K.}~\bibnamefont {Saaidi}}, \
  and\ \bibinfo {author} {\bibfnamefont {T.}~\bibnamefont {Golanbari}},\ }\href
  {\doibase 10.1103/PhysRevD.97.083006} {\bibfield  {journal} {\bibinfo
  {journal} {Phys. Rev.}\ }\textbf {\bibinfo {volume} {D97}},\ \bibinfo {pages}
  {083006} (\bibinfo {year} {2018}{\natexlab{a}})},\ \Eprint
  {http://arxiv.org/abs/1801.03487} {arXiv:1801.03487 [hep-ph]} \BibitemShut
  {NoStop}%
\bibitem [{\citenamefont {Gao}\ \emph {et~al.}(2018)\citenamefont {Gao},
  \citenamefont {Gong},\ and\ \citenamefont {Fei}}]{Gao:2018tdb}%
  \BibitemOpen
  \bibfield  {author} {\bibinfo {author} {\bibfnamefont {Q.}~\bibnamefont
  {Gao}}, \bibinfo {author} {\bibfnamefont {Y.}~\bibnamefont {Gong}}, \ and\
  \bibinfo {author} {\bibfnamefont {Q.}~\bibnamefont {Fei}},\ }\href {\doibase
  10.1088/1475-7516/2018/05/005} {\bibfield  {journal} {\bibinfo  {journal}
  {JCAP}\ }\textbf {\bibinfo {volume} {1805}},\ \bibinfo {pages} {005}
  (\bibinfo {year} {2018})},\ \Eprint {http://arxiv.org/abs/1801.09208}
  {arXiv:1801.09208 [gr-qc]} \BibitemShut {NoStop}%
\bibitem [{\citenamefont {Gao}(2018)}]{Gao:2018cpp}%
  \BibitemOpen
  \bibfield  {author} {\bibinfo {author} {\bibfnamefont {Q.}~\bibnamefont
  {Gao}},\ }\href {\doibase 10.1007/s11433-018-9197-2} {\bibfield  {journal}
  {\bibinfo  {journal} {Sci. China Phys. Mech. Astron.}\ }\textbf {\bibinfo
  {volume} {61}},\ \bibinfo {pages} {070411} (\bibinfo {year} {2018})},\
  \Eprint {http://arxiv.org/abs/1802.01986} {arXiv:1802.01986 [gr-qc]}
  \BibitemShut {NoStop}%
\bibitem [{\citenamefont {Anguelova}\ \emph {et~al.}(2017)\citenamefont
  {Anguelova}, \citenamefont {Suranyi},\ and\ \citenamefont
  {Rohana~Wijewardhana}}]{Anguelova:2018ntr}%
  \BibitemOpen
  \bibfield  {author} {\bibinfo {author} {\bibfnamefont {L.}~\bibnamefont
  {Anguelova}}, \bibinfo {author} {\bibfnamefont {P.}~\bibnamefont {Suranyi}},
  \ and\ \bibinfo {author} {\bibfnamefont {L.~C.}\ \bibnamefont
  {Rohana~Wijewardhana}},\ }\bibfield  {booktitle} {\emph {\bibinfo {booktitle}
  {{Proceedings, 10th International Symposium on Quantum theory and symmetries
  (QTS10) and 12th International Workshop on Lie Theory and Its Applications in
  Physics (LT12): Varna, Bulgaria, June 19-25, 2017}}},\ }\href {\doibase
  10.1007/978-981-13-2179-5_11} {\bibfield  {journal} {\bibinfo  {journal}
  {Springer Proc. Math. Stat.}\ }\textbf {\bibinfo {volume} {255}},\ \bibinfo
  {pages} {161} (\bibinfo {year} {2017})},\ \Eprint
  {http://arxiv.org/abs/1802.02625} {arXiv:1802.02625 [hep-th]} \BibitemShut
  {NoStop}%
\bibitem [{\citenamefont {Mohammadi}\ and\ \citenamefont
  {Saaidi}(2018)}]{Mohammadi:2018wfk}%
  \BibitemOpen
  \bibfield  {author} {\bibinfo {author} {\bibfnamefont {A.}~\bibnamefont
  {Mohammadi}}\ and\ \bibinfo {author} {\bibfnamefont {K.}~\bibnamefont
  {Saaidi}},\ }\href@noop {} {\  (\bibinfo {year} {2018})},\ \Eprint
  {http://arxiv.org/abs/1803.01715} {arXiv:1803.01715 [astro-ph.CO]}
  \BibitemShut {NoStop}%
\bibitem [{\citenamefont {Karam}\ \emph {et~al.}(2018)\citenamefont {Karam},
  \citenamefont {Marzola}, \citenamefont {Pappas}, \citenamefont {Racioppi},\
  and\ \citenamefont {Tamvakis}}]{Karam:2017rpw}%
  \BibitemOpen
  \bibfield  {author} {\bibinfo {author} {\bibfnamefont {A.}~\bibnamefont
  {Karam}}, \bibinfo {author} {\bibfnamefont {L.}~\bibnamefont {Marzola}},
  \bibinfo {author} {\bibfnamefont {T.}~\bibnamefont {Pappas}}, \bibinfo
  {author} {\bibfnamefont {A.}~\bibnamefont {Racioppi}}, \ and\ \bibinfo
  {author} {\bibfnamefont {K.}~\bibnamefont {Tamvakis}},\ }\href {\doibase
  10.1088/1475-7516/2018/05/011} {\bibfield  {journal} {\bibinfo  {journal}
  {JCAP}\ }\textbf {\bibinfo {volume} {1805}},\ \bibinfo {pages} {011}
  (\bibinfo {year} {2018})},\ \Eprint {http://arxiv.org/abs/1711.09861}
  {arXiv:1711.09861 [astro-ph.CO]} \BibitemShut {NoStop}%
\bibitem [{\citenamefont {Morse}\ and\ \citenamefont
  {Kinney}(2018)}]{Morse:2018kda}%
  \BibitemOpen
  \bibfield  {author} {\bibinfo {author} {\bibfnamefont {M.~J.~P.}\
  \bibnamefont {Morse}}\ and\ \bibinfo {author} {\bibfnamefont {W.~H.}\
  \bibnamefont {Kinney}},\ }\href {\doibase 10.1103/PhysRevD.97.123519}
  {\bibfield  {journal} {\bibinfo  {journal} {Phys. Rev.}\ }\textbf {\bibinfo
  {volume} {D97}},\ \bibinfo {pages} {123519} (\bibinfo {year} {2018})},\
  \Eprint {http://arxiv.org/abs/1804.01927} {arXiv:1804.01927 [astro-ph.CO]}
  \BibitemShut {NoStop}%
\bibitem [{\citenamefont {Cruces}\ \emph {et~al.}(2019)\citenamefont {Cruces},
  \citenamefont {Germani},\ and\ \citenamefont {Prokopec}}]{Cruces:2018cvq}%
  \BibitemOpen
  \bibfield  {author} {\bibinfo {author} {\bibfnamefont {D.}~\bibnamefont
  {Cruces}}, \bibinfo {author} {\bibfnamefont {C.}~\bibnamefont {Germani}}, \
  and\ \bibinfo {author} {\bibfnamefont {T.}~\bibnamefont {Prokopec}},\ }\href
  {\doibase 10.1088/1475-7516/2019/03/048} {\bibfield  {journal} {\bibinfo
  {journal} {JCAP}\ }\textbf {\bibinfo {volume} {1903}},\ \bibinfo {pages}
  {048} (\bibinfo {year} {2019})},\ \Eprint {http://arxiv.org/abs/1807.09057}
  {arXiv:1807.09057 [gr-qc]} \BibitemShut {NoStop}%
\bibitem [{\citenamefont {Mohammadi}\ \emph
  {et~al.}(2018{\natexlab{b}})\citenamefont {Mohammadi}, \citenamefont
  {Saaidi},\ and\ \citenamefont {Golanbari}}]{Mohammadi:2018zkf}%
  \BibitemOpen
  \bibfield  {author} {\bibinfo {author} {\bibfnamefont {A.}~\bibnamefont
  {Mohammadi}}, \bibinfo {author} {\bibfnamefont {K.}~\bibnamefont {Saaidi}}, \
  and\ \bibinfo {author} {\bibfnamefont {T.}~\bibnamefont {Golanbari}},\
  }\href@noop {} {\  (\bibinfo {year} {2018}{\natexlab{b}})},\ \Eprint
  {http://arxiv.org/abs/1808.07246} {arXiv:1808.07246 [gr-qc]} \BibitemShut
  {NoStop}%
\bibitem [{\citenamefont {Boisseau}\ and\ \citenamefont
  {Giacomini}(2018)}]{Boisseau:2018rgy}%
  \BibitemOpen
  \bibfield  {author} {\bibinfo {author} {\bibfnamefont {B.}~\bibnamefont
  {Boisseau}}\ and\ \bibinfo {author} {\bibfnamefont {H.}~\bibnamefont
  {Giacomini}},\ }\href@noop {} {\  (\bibinfo {year} {2018})},\ \Eprint
  {http://arxiv.org/abs/1809.09169} {arXiv:1809.09169 [gr-qc]} \BibitemShut
  {NoStop}%
\bibitem [{\citenamefont {Firouzjahi}\ \emph {et~al.}(2019)\citenamefont
  {Firouzjahi}, \citenamefont {Nassiri-Rad},\ and\ \citenamefont
  {Noorbala}}]{Firouzjahi:2018vet}%
  \BibitemOpen
  \bibfield  {author} {\bibinfo {author} {\bibfnamefont {H.}~\bibnamefont
  {Firouzjahi}}, \bibinfo {author} {\bibfnamefont {A.}~\bibnamefont
  {Nassiri-Rad}}, \ and\ \bibinfo {author} {\bibfnamefont {M.}~\bibnamefont
  {Noorbala}},\ }\href {\doibase 10.1088/1475-7516/2019/01/040} {\bibfield
  {journal} {\bibinfo  {journal} {JCAP}\ }\textbf {\bibinfo {volume} {1901}},\
  \bibinfo {pages} {040} (\bibinfo {year} {2019})},\ \Eprint
  {http://arxiv.org/abs/1811.02175} {arXiv:1811.02175 [hep-th]} \BibitemShut
  {NoStop}%
\bibitem [{\citenamefont {Matarrese}\ \emph {et~al.}(2019)\citenamefont
  {Matarrese}, \citenamefont {Pilo},\ and\ \citenamefont
  {Rollo}}]{Matarrese:2018qqo}%
  \BibitemOpen
  \bibfield  {author} {\bibinfo {author} {\bibfnamefont {S.}~\bibnamefont
  {Matarrese}}, \bibinfo {author} {\bibfnamefont {L.}~\bibnamefont {Pilo}}, \
  and\ \bibinfo {author} {\bibfnamefont {R.}~\bibnamefont {Rollo}},\ }\href
  {\doibase 10.1088/1475-7516/2019/04/017} {\bibfield  {journal} {\bibinfo
  {journal} {JCAP}\ }\textbf {\bibinfo {volume} {1904}},\ \bibinfo {pages}
  {017} (\bibinfo {year} {2019})},\ \Eprint {http://arxiv.org/abs/1812.03844}
  {arXiv:1812.03844 [gr-qc]} \BibitemShut {NoStop}%
\bibitem [{\citenamefont {Pattison}\ \emph {et~al.}(2018)\citenamefont
  {Pattison}, \citenamefont {Vennin}, \citenamefont {Assadullahi},\ and\
  \citenamefont {Wands}}]{Pattison:2018bct}%
  \BibitemOpen
  \bibfield  {author} {\bibinfo {author} {\bibfnamefont {C.}~\bibnamefont
  {Pattison}}, \bibinfo {author} {\bibfnamefont {V.}~\bibnamefont {Vennin}},
  \bibinfo {author} {\bibfnamefont {H.}~\bibnamefont {Assadullahi}}, \ and\
  \bibinfo {author} {\bibfnamefont {D.}~\bibnamefont {Wands}},\ }\href
  {\doibase 10.1088/1475-7516/2018/08/048} {\bibfield  {journal} {\bibinfo
  {journal} {JCAP}\ }\textbf {\bibinfo {volume} {1808}},\ \bibinfo {pages}
  {048} (\bibinfo {year} {2018})},\ \Eprint {http://arxiv.org/abs/1806.09553}
  {arXiv:1806.09553 [astro-ph.CO]} \BibitemShut {NoStop}%
\bibitem [{\citenamefont {Ozsoy}\ \emph {et~al.}(2019)\citenamefont {Ozsoy},
  \citenamefont {Mylova}, \citenamefont {Parameswaran}, \citenamefont {Powell},
  \citenamefont {Tasinato},\ and\ \citenamefont {Zavala}}]{Ozsoy:2019slf}%
  \BibitemOpen
  \bibfield  {author} {\bibinfo {author} {\bibfnamefont {O.}~\bibnamefont
  {Ozsoy}}, \bibinfo {author} {\bibfnamefont {M.}~\bibnamefont {Mylova}},
  \bibinfo {author} {\bibfnamefont {S.}~\bibnamefont {Parameswaran}}, \bibinfo
  {author} {\bibfnamefont {C.}~\bibnamefont {Powell}}, \bibinfo {author}
  {\bibfnamefont {G.}~\bibnamefont {Tasinato}}, \ and\ \bibinfo {author}
  {\bibfnamefont {I.}~\bibnamefont {Zavala}},\ }\href@noop {} {\  (\bibinfo
  {year} {2019})},\ \Eprint {http://arxiv.org/abs/1902.04976} {arXiv:1902.04976
  [hep-th]} \BibitemShut {NoStop}%
\bibitem [{\citenamefont {León}\ \emph {et~al.}(2019)\citenamefont {León},
  \citenamefont {Pujol}, \citenamefont {Landau},\ and\ \citenamefont
  {Piccirilli}}]{Leon:2019jsl}%
  \BibitemOpen
  \bibfield  {author} {\bibinfo {author} {\bibfnamefont {G.}~\bibnamefont
  {León}}, \bibinfo {author} {\bibfnamefont {A.}~\bibnamefont {Pujol}},
  \bibinfo {author} {\bibfnamefont {S.~J.}\ \bibnamefont {Landau}}, \ and\
  \bibinfo {author} {\bibfnamefont {M.~P.}\ \bibnamefont {Piccirilli}},\ }\href
  {\doibase 10.1016/j.dark.2019.100285} {\bibfield  {journal} {\bibinfo
  {journal} {Phys. Dark Univ.}\ ,\ \bibinfo {pages} {100285}} (\bibinfo {year}
  {2019})},\ \Eprint {http://arxiv.org/abs/1902.08696} {arXiv:1902.08696
  [astro-ph.CO]} \BibitemShut {NoStop}%
\bibitem [{\citenamefont {Gao}\ \emph {et~al.}(2019)\citenamefont {Gao},
  \citenamefont {Gong},\ and\ \citenamefont {Yi}}]{Gao:2019sbz}%
  \BibitemOpen
  \bibfield  {author} {\bibinfo {author} {\bibfnamefont {Q.}~\bibnamefont
  {Gao}}, \bibinfo {author} {\bibfnamefont {Y.}~\bibnamefont {Gong}}, \ and\
  \bibinfo {author} {\bibfnamefont {Z.}~\bibnamefont {Yi}},\ }\href@noop {} {\
  (\bibinfo {year} {2019})},\ \Eprint {http://arxiv.org/abs/1901.04646}
  {arXiv:1901.04646 [gr-qc]} \BibitemShut {NoStop}%
\bibitem [{\citenamefont {Nicholson}\ and\ \citenamefont
  {Contaldi}(2008)}]{Nicholson:2007by}%
  \BibitemOpen
  \bibfield  {author} {\bibinfo {author} {\bibfnamefont {G.}~\bibnamefont
  {Nicholson}}\ and\ \bibinfo {author} {\bibfnamefont {C.~R.}\ \bibnamefont
  {Contaldi}},\ }\href {\doibase 10.1088/1475-7516/2008/01/002} {\bibfield
  {journal} {\bibinfo  {journal} {JCAP}\ }\textbf {\bibinfo {volume} {0801}},\
  \bibinfo {pages} {002} (\bibinfo {year} {2008})},\ \Eprint
  {http://arxiv.org/abs/astro-ph/0701783} {arXiv:astro-ph/0701783 [astro-ph]}
  \BibitemShut {NoStop}%
\bibitem [{\citenamefont {Contaldi}\ \emph {et~al.}(2003)\citenamefont
  {Contaldi}, \citenamefont {Peloso}, \citenamefont {Kofman},\ and\
  \citenamefont {Linde}}]{Contaldi:2003zv}%
  \BibitemOpen
  \bibfield  {author} {\bibinfo {author} {\bibfnamefont {C.~R.}\ \bibnamefont
  {Contaldi}}, \bibinfo {author} {\bibfnamefont {M.}~\bibnamefont {Peloso}},
  \bibinfo {author} {\bibfnamefont {L.}~\bibnamefont {Kofman}}, \ and\ \bibinfo
  {author} {\bibfnamefont {A.~D.}\ \bibnamefont {Linde}},\ }\href {\doibase
  10.1088/1475-7516/2003/07/002} {\bibfield  {journal} {\bibinfo  {journal}
  {JCAP}\ }\textbf {\bibinfo {volume} {0307}},\ \bibinfo {pages} {002}
  (\bibinfo {year} {2003})},\ \Eprint {http://arxiv.org/abs/astro-ph/0303636}
  {arXiv:astro-ph/0303636 [astro-ph]} \BibitemShut {NoStop}%
\bibitem [{\citenamefont {Starobinsky}(1980)}]{Starobinsky:1980te}%
  \BibitemOpen
  \bibfield  {author} {\bibinfo {author} {\bibfnamefont {A.~A.}\ \bibnamefont
  {Starobinsky}},\ }\href {\doibase 10.1016/0370-2693(80)90670-X} {\bibfield
  {journal} {\bibinfo  {journal} {Phys. Lett.}\ }\textbf {\bibinfo {volume}
  {B91}},\ \bibinfo {pages} {99} (\bibinfo {year} {1980})},\ \bibinfo {note}
  {[,771(1980)]}\BibitemShut {NoStop}%
\bibitem [{\citenamefont {Sato}(1981{\natexlab{a}})}]{Sato:1981ds}%
  \BibitemOpen
  \bibfield  {author} {\bibinfo {author} {\bibfnamefont {K.}~\bibnamefont
  {Sato}},\ }\href {\doibase 10.1016/0370-2693(81)90805-4} {\bibfield
  {journal} {\bibinfo  {journal} {Phys. Lett.}\ }\textbf {\bibinfo {volume}
  {99B}},\ \bibinfo {pages} {66} (\bibinfo {year} {1981}{\natexlab{a}})},\
  \bibinfo {note} {[Adv. Ser. Astrophys. Cosmol.3,134(1987)]}\BibitemShut
  {NoStop}%
\bibitem [{\citenamefont {Sato}(1981{\natexlab{b}})}]{Sato:1980yn}%
  \BibitemOpen
  \bibfield  {author} {\bibinfo {author} {\bibfnamefont {K.}~\bibnamefont
  {Sato}},\ }\href@noop {} {\bibfield  {journal} {\bibinfo  {journal} {Mon.
  Not. Roy. Astron. Soc.}\ }\textbf {\bibinfo {volume} {195}},\ \bibinfo
  {pages} {467} (\bibinfo {year} {1981}{\natexlab{b}})}\BibitemShut {NoStop}%
\bibitem [{\citenamefont {Kazanas}(1980)}]{Kazanas:1980tx}%
  \BibitemOpen
  \bibfield  {author} {\bibinfo {author} {\bibfnamefont {D.}~\bibnamefont
  {Kazanas}},\ }\href {\doibase 10.1086/183361} {\bibfield  {journal} {\bibinfo
   {journal} {Astrophys. J.}\ }\textbf {\bibinfo {volume} {241}},\ \bibinfo
  {pages} {L59} (\bibinfo {year} {1980})}\BibitemShut {NoStop}%
\bibitem [{\citenamefont {Guth}(1981)}]{Guth:1980zm}%
  \BibitemOpen
  \bibfield  {author} {\bibinfo {author} {\bibfnamefont {A.~H.}\ \bibnamefont
  {Guth}},\ }\href {\doibase 10.1103/PhysRevD.23.347} {\bibfield  {journal}
  {\bibinfo  {journal} {Phys. Rev.}\ }\textbf {\bibinfo {volume} {D23}},\
  \bibinfo {pages} {347} (\bibinfo {year} {1981})},\ \bibinfo {note} {[Adv.
  Ser. Astrophys. Cosmol.3,139(1987)]}\BibitemShut {NoStop}%
\bibitem [{\citenamefont {Linde}(1982)}]{Linde:1981mu}%
  \BibitemOpen
  \bibfield  {author} {\bibinfo {author} {\bibfnamefont {A.~D.}\ \bibnamefont
  {Linde}},\ }\bibfield  {booktitle} {\emph {\bibinfo {booktitle} {{QUANTUM
  COSMOLOGY}}},\ }\href {\doibase 10.1016/0370-2693(82)91219-9} {\bibfield
  {journal} {\bibinfo  {journal} {Phys. Lett.}\ }\textbf {\bibinfo {volume}
  {108B}},\ \bibinfo {pages} {389} (\bibinfo {year} {1982})},\ \bibinfo {note}
  {[Adv. Ser. Astrophys. Cosmol.3,149(1987)]}\BibitemShut {NoStop}%
\bibitem [{\citenamefont {Albrecht}\ and\ \citenamefont
  {Steinhardt}(1982)}]{Albrecht:1982wi}%
  \BibitemOpen
  \bibfield  {author} {\bibinfo {author} {\bibfnamefont {A.}~\bibnamefont
  {Albrecht}}\ and\ \bibinfo {author} {\bibfnamefont {P.~J.}\ \bibnamefont
  {Steinhardt}},\ }\href {\doibase 10.1103/PhysRevLett.48.1220} {\bibfield
  {journal} {\bibinfo  {journal} {Phys. Rev. Lett.}\ }\textbf {\bibinfo
  {volume} {48}},\ \bibinfo {pages} {1220} (\bibinfo {year} {1982})},\ \bibinfo
  {note} {[Adv. Ser. Astrophys. Cosmol.3,158(1987)]}\BibitemShut {NoStop}%
\bibitem [{\citenamefont {Muslimov}(1990)}]{Muslimov:1990be}%
  \BibitemOpen
  \bibfield  {author} {\bibinfo {author} {\bibfnamefont {A.~G.}\ \bibnamefont
  {Muslimov}},\ }\href {\doibase 10.1088/0264-9381/7/2/015} {\bibfield
  {journal} {\bibinfo  {journal} {Class. Quant. Grav.}\ }\textbf {\bibinfo
  {volume} {7}},\ \bibinfo {pages} {231} (\bibinfo {year} {1990})}\BibitemShut
  {NoStop}%
\bibitem [{\citenamefont {Salopek}\ and\ \citenamefont
  {Bond}(1990)}]{Salopek:1990jq}%
  \BibitemOpen
  \bibfield  {author} {\bibinfo {author} {\bibfnamefont {D.~S.}\ \bibnamefont
  {Salopek}}\ and\ \bibinfo {author} {\bibfnamefont {J.~R.}\ \bibnamefont
  {Bond}},\ }\href {\doibase 10.1103/PhysRevD.42.3936} {\bibfield  {journal}
  {\bibinfo  {journal} {Phys. Rev.}\ }\textbf {\bibinfo {volume} {D42}},\
  \bibinfo {pages} {3936} (\bibinfo {year} {1990})}\BibitemShut {NoStop}%
\bibitem [{\citenamefont {Lidsey}\ \emph {et~al.}(1997)\citenamefont {Lidsey},
  \citenamefont {Liddle}, \citenamefont {Kolb}, \citenamefont {Copeland},
  \citenamefont {Barreiro},\ and\ \citenamefont {Abney}}]{Lidsey:1995np}%
  \BibitemOpen
  \bibfield  {author} {\bibinfo {author} {\bibfnamefont {J.~E.}\ \bibnamefont
  {Lidsey}}, \bibinfo {author} {\bibfnamefont {A.~R.}\ \bibnamefont {Liddle}},
  \bibinfo {author} {\bibfnamefont {E.~W.}\ \bibnamefont {Kolb}}, \bibinfo
  {author} {\bibfnamefont {E.~J.}\ \bibnamefont {Copeland}}, \bibinfo {author}
  {\bibfnamefont {T.}~\bibnamefont {Barreiro}}, \ and\ \bibinfo {author}
  {\bibfnamefont {M.}~\bibnamefont {Abney}},\ }\href {\doibase
  10.1103/RevModPhys.69.373} {\bibfield  {journal} {\bibinfo  {journal} {Rev.
  Mod. Phys.}\ }\textbf {\bibinfo {volume} {69}},\ \bibinfo {pages} {373}
  (\bibinfo {year} {1997})},\ \Eprint {http://arxiv.org/abs/astro-ph/9508078}
  {arXiv:astro-ph/9508078 [astro-ph]} \BibitemShut {NoStop}%
\bibitem [{\citenamefont {Passaglia}\ \emph {et~al.}(2019)\citenamefont
  {Passaglia}, \citenamefont {Hu},\ and\ \citenamefont
  {Motohashi}}]{Passaglia:2018ixg}%
  \BibitemOpen
  \bibfield  {author} {\bibinfo {author} {\bibfnamefont {S.}~\bibnamefont
  {Passaglia}}, \bibinfo {author} {\bibfnamefont {W.}~\bibnamefont {Hu}}, \
  and\ \bibinfo {author} {\bibfnamefont {H.}~\bibnamefont {Motohashi}},\ }\href
  {\doibase 10.1103/PhysRevD.99.043536} {\bibfield  {journal} {\bibinfo
  {journal} {Phys. Rev.}\ }\textbf {\bibinfo {volume} {D99}},\ \bibinfo {pages}
  {043536} (\bibinfo {year} {2019})},\ \Eprint
  {http://arxiv.org/abs/1812.08243} {arXiv:1812.08243 [astro-ph.CO]}
  \BibitemShut {NoStop}%
\bibitem [{\citenamefont {Morse}(2019)}]{michael_j_p_morse_2019_3066362}%
  \BibitemOpen
  \bibfield  {author} {\bibinfo {author} {\bibfnamefont {M.~J.~P.}\
  \bibnamefont {Morse}},\ }\href {\doibase 10.5281/zenodo.3066362} {\enquote
  {\bibinfo {title} {Kinneyresearch/constantrollrelease: Public release},}\ }
  (\bibinfo {year} {2019})\BibitemShut {NoStop}%
\bibitem [{\citenamefont {Linde}(1994)}]{Linde:1993cn}%
  \BibitemOpen
  \bibfield  {author} {\bibinfo {author} {\bibfnamefont {A.~D.}\ \bibnamefont
  {Linde}},\ }\href {\doibase 10.1103/PhysRevD.49.748} {\bibfield  {journal}
  {\bibinfo  {journal} {Phys. Rev.}\ }\textbf {\bibinfo {volume} {D49}},\
  \bibinfo {pages} {748} (\bibinfo {year} {1994})},\ \Eprint
  {http://arxiv.org/abs/astro-ph/9307002} {arXiv:astro-ph/9307002 [astro-ph]}
  \BibitemShut {NoStop}%
\bibitem [{\citenamefont {Akrami}\ \emph {et~al.}(2018)\citenamefont {Akrami}
  \emph {et~al.}}]{Akrami:2018odb}%
  \BibitemOpen
  \bibfield  {author} {\bibinfo {author} {\bibfnamefont {Y.}~\bibnamefont
  {Akrami}} \emph {et~al.} (\bibinfo {collaboration} {Planck}),\ }\href@noop {}
  {\  (\bibinfo {year} {2018})},\ \Eprint {http://arxiv.org/abs/1807.06211}
  {arXiv:1807.06211 [astro-ph.CO]} \BibitemShut {NoStop}%
\bibitem [{\citenamefont {Ade}\ \emph {et~al.}(2018)\citenamefont {Ade} \emph
  {et~al.}}]{Ade:2018gkx}%
  \BibitemOpen
  \bibfield  {author} {\bibinfo {author} {\bibfnamefont {P.~A.~R.}\
  \bibnamefont {Ade}} \emph {et~al.} (\bibinfo {collaboration} {BICEP2, Keck
  Array}),\ }\href {\doibase 10.1103/PhysRevLett.121.221301} {\bibfield
  {journal} {\bibinfo  {journal} {Phys. Rev. Lett.}\ }\textbf {\bibinfo
  {volume} {121}},\ \bibinfo {pages} {221301} (\bibinfo {year} {2018})},\
  \Eprint {http://arxiv.org/abs/1810.05216} {arXiv:1810.05216 [astro-ph.CO]}
  \BibitemShut {NoStop}%
\bibitem [{\citenamefont {Tzirakis}\ and\ \citenamefont
  {Kinney}(2007)}]{Tzirakis:2007bf}%
  \BibitemOpen
  \bibfield  {author} {\bibinfo {author} {\bibfnamefont {K.}~\bibnamefont
  {Tzirakis}}\ and\ \bibinfo {author} {\bibfnamefont {W.~H.}\ \bibnamefont
  {Kinney}},\ }\href {\doibase 10.1103/PhysRevD.75.123510} {\bibfield
  {journal} {\bibinfo  {journal} {Phys. Rev.}\ }\textbf {\bibinfo {volume}
  {D75}},\ \bibinfo {pages} {123510} (\bibinfo {year} {2007})},\ \Eprint
  {http://arxiv.org/abs/astro-ph/0701432} {arXiv:astro-ph/0701432 [astro-ph]}
  \BibitemShut {NoStop}%
\bibitem [{\citenamefont {Kinney}(2019)}]{Kinney:2018kew}%
  \BibitemOpen
  \bibfield  {author} {\bibinfo {author} {\bibfnamefont {W.~H.}\ \bibnamefont
  {Kinney}},\ }\href {\doibase 10.1103/PhysRevLett.122.081302} {\bibfield
  {journal} {\bibinfo  {journal} {Phys. Rev. Lett.}\ }\textbf {\bibinfo
  {volume} {122}},\ \bibinfo {pages} {081302} (\bibinfo {year} {2019})},\
  \Eprint {http://arxiv.org/abs/1811.11698} {arXiv:1811.11698 [astro-ph.CO]}
  \BibitemShut {NoStop}%
\bibitem [{\citenamefont {Jain}\ and\ \citenamefont
  {Hertzberg}(2019)}]{Jain:2019gsq}%
  \BibitemOpen
  \bibfield  {author} {\bibinfo {author} {\bibfnamefont {M.}~\bibnamefont
  {Jain}}\ and\ \bibinfo {author} {\bibfnamefont {M.~P.}\ \bibnamefont
  {Hertzberg}},\ }\href@noop {} {\  (\bibinfo {year} {2019})},\ \Eprint
  {http://arxiv.org/abs/1904.04262} {arXiv:1904.04262 [astro-ph.CO]}
  \BibitemShut {NoStop}%
\end{thebibliography}%

\end{document}